\documentclass{biophys-new}

\usepackage{verbatim}
\usepackage{mathtools}
\usepackage{enumitem}
\usepackage{siunitx}
\usepackage{tikz}

\usepackage{xr}
\newcommand{\T}{\ensuremath{\mathrm{T}}}
\newcommand{\e}{\ensuremath{\mathrm{e}}}
\newcommand{\bd}[1]{\ensuremath{\boldsymbol{#1}}}

\newcommand{\dd}{\ensuremath{\mathrm{d}}}
\newcommand{\da}{\ensuremath{_{\mathrm{(D,\alpha)}}}}
\DeclareMathOperator*{\E}{\mathbb{E}}

\DeclareMathOperator*{\var}{{var}}
\DeclareMathOperator*{\cov}{{cov}}
\newcommand{\review}[1]{{{#1}}}

\usepackage[colorlinks,allcolors=cyan!70!black]{hyperref}

\title{Improved mean squared displacement analysis for anomalous single
particle trajectories}
\runningtitle{Improved mean squared displacement analysis for anomalous single	particle trajectories}

\author[1,*]{Jakub \'Sl\k{e}zak}
\author[1]{Joanna Janczura}
\author[2]{Diego Krapf}
\author[3,4]{Ralf Metzler}
\runningauthor{J. \'Sl\k{e}zak, J. Janczura, D. Krapf, R. Metzler } 

\affil[1]{Hugo Steinhaus Center,Wroc{\l}aw University of Science and Technology, Wrocław 58-330, Poland}
\affil[2]{School of Biomedical and Chemical Engineering, Colorado	State University, Fort Collins, CO 80523, USA}
\affil[3]{University of Potsdam, Institute of Physics \& Astronomy Potsdam 14476, Germany}
\affil[4]{Asia Pacific Center for Theoretical Physics, Pohang 37673, Republic of	Korea}

\corrauthor[*]{jakub.slezak@pwr.edu.pl}

\papertype{Article}

\begin{document}

\begin{frontmatter}
	
	\begin{abstract}
		The mean squared displacement (MSD) is a cornerstone in the analysis
		of diffusion processes in complex media. When the system is heterogeneous
		and, in particular, when single-particle trajectories are short,
		it is essential to extract maximal information from each measured
		trajectory. This is typically done by time-averaging squared increments and examining the 
		scaling of the time-averaged MSD in log–log space.
		However, classical regression methods perform poorly in this setting
		because time-averaging introduces correlations aggravated by those inherent to
		anomalous diffusion. We tackle these limitations by applying a generalized least squares
		framework, which substantially reduces
		variance and bias in diffusion parameter estimates, especially for short
		($\approx100$ points) and ultra-short ($\approx10$ points) trajectories. The
		method is fully automated and requires no supervision. 
		Furthermore, it enables prediction of estimation error probability density, which is asymptotically Gaussian, for both classical and enhanced approaches. Leveraging this prediction, we introduce a specialized deconvolution algorithm that reconstructs the underlying
		particle ensemble structure from experimental data.
	\end{abstract}
	
	\begin{sigstatement}
		Statistically analysing the mean squared displacement is a fundamental task in understanding multiple facets of molecular kinetics. Due to experimental limitations the considered trajectories are often quite short which causes errors to significantly affect the estimated population structure of studied particles. Under these common conditions the proposed method for decreasing these errors and predicting their amplitude can directly increase the control of the experimentalist. It is also a mathematically necessary step to unambiguously answer questions like if there are distinct subpopulations of particles in the data or are the observed features only statistical artefacts. The study shows the method works for simulations which imitate experimental measurements and yields new information for real anomalous diffusion data.
	\end{sigstatement}
\end{frontmatter}

\section*{Introduction}

Single particle tracking (SPT) has emerged as a vital tool for uncovering the
dynamic properties of biological and other complex systems
\cite{levi2007exploring,hofling2013anomalous,barkai2012strange,manzo2015review,
shen2017single,norregaard2017manipulation,krapf2019strange,muozGil2021,andi2}.
By tracking individual particles, it is possible to explore
material and chemical properties within cellular environments, including the
cytoplasm \cite{caspi2000enhanced,sabri2020elucidating}, nucleus
\cite{bronstein2009transient,lucas20143d}, plasma membrane
\cite{weigel2011ergodic,manzo2015weak}, and organelles
\cite{holcman2018single,obara2024motion}. These studies illuminate critical
processes such as protein interactions \cite{low2011erbb1}, cell signaling
\cite{choquet2003role}, chromatin remodeling \cite{dipierro2018}, genetic
regulation \cite{moon2019multicolour,saxton2023live}, viral entry into cells \cite{brandenburg2007virus,van2008dissecting}, and endocytosis
\cite{jin2008single,weigel2013quantifying,fox2013plasma}. Beyond intracellular
dynamics, the use of particle tracking extends to other diverse fields,
such as the migratory behavior of animals \cite{animalBigData}, cellular
motility \cite{berg2000motile}, colloidal cargo transport on cell carpets
\cite{beta,grossmann2024}, human mobility \cite{brockmann}, and the study of
percolation phenomena in porous materials \cite{edery2014origins,
wu2020nanoparticle}. This versatility highlights the breadth of SPT in
advancing both fundamental and applied research.

Central to the analysis of trajectories obtained from SPT is the calculation
of the mean squared displacement (MSD), a metric extensively employed to
quantify the diffusion of tracers \cite{barkai2012strange,hofling2013anomalous,
manzo2015review,norregaard2017manipulation,kepten2015guide}. The classical MSD is
defined as the ensemble average over all trajectories of the squares of the
displacements, $\langle\delta^2(t)\rangle=\big\langle\big(\boldsymbol{R}(t)-
\boldsymbol{R}(0)\big)^2\big\rangle$, at a given time $t$. We will refer to
it as an ensemble-averaged MSD (EA-MSD). However, for individual trajectories, 
this measure is impossible to calculate, and one uses the alternative
time-averaged MSD (TA-MSD), which is a sliding average along the time series $\boldsymbol{R}(t)$ \cite{barkai2012strange,manzo2015review}, 
\begin{equation}
\label{eq:tamsdDef}
\overline{\delta^2(t)}\coloneqq\frac{1}{T-t}\int_0^{T-t}\dd\tau\big(\boldsymbol{
R}(\tau+t)-\boldsymbol{R}(\tau)\big)^2 ,
\end{equation}
where $T$ denotes the measurement time. Averaging the
TA-MSD over $N$ trajectories, provides the mean
TA-MSD $\big\langle\overline{\delta^2(t)}\big\rangle=N^{-1}\sum_{i=1}^N\overline{
\delta_i^2(t)}$. The two complementary MSD definitions provide insights into
the nature of the observed particle motion and its governing principles,
rendering the MSD a cornerstone of SPT-based studies.

For a system of identical particles performing Brownian motion, the EA-MSD
scales linearly with time, $\langle\delta^2(t)\rangle=2dDt$, where $D$ is the
diffusion coefficient and $d$ the dimension of the problem being studied.
However, many complex systems deviate from the classical Brownian behavior,
exhibiting anomalous diffusion characterized by a non-linear relation between
the EA-MSD and time \cite{barkai2012strange,hofling2013anomalous,pccp}. In such
cases, the EA-MSD often displays a power-law behavior, $\langle\delta^2(t)\rangle= 2dD
t^\alpha$, where the anomalous exponent $\alpha$ describes the
type of anomalous diffusion. When $\alpha=1$, the system is normal-diffusive, but $\alpha>1$ indicates superdiffusion and $0<\alpha<1$
denotes subdiffusion. Another special case is $\alpha=2$, which represents
ballistic motion. In these cases, the coefficient $D$ has the units of $\mathrm{length}^2/\mathrm{time}^{\alpha}$.
An accurate estimation of $\alpha$ is necessary to identify the type of
anomalous diffusion at hand and only then, proper models can be used to understand
the underlying physical properties of the system. In particular, $\alpha$ is
close to 1 in many practical situations, leading to erroneous claims of
Brownian or anomalous diffusion. Thus reliable ways to decipher the precise
value of $\alpha$ are needed \cite{sposini2022towards,muozGil2021,andi2}.

Depending on the physical nature of the system, the diffusive motion of tracer
particles may be described by a diverse range of models \cite{pccp,igorsoft,
krapf2015mechanisms}, including the continuous-time random walk (CTRW)
\cite{scher1975anomalous,hughes}, fractional Brownian motion (FBM)
\cite{mandelbrot1,kolmogorov}, annealed transient time motion (ATTM)
\cite{lapeyre}, fractional Langevin equation \cite{kouPRL}, diffusion in
fractal environments \cite{o1985diffusion,sadegh2017plasma}, or active Brownian motion
\cite{joo}. A key distinction between these processes lies in whether the
statistics of their increments are constant over time, i.e., if these increments are
stationary \cite{igoryasmin}. If they are not, the EA-MSD and TA-MSD
generally differ, and the diffusion can be anomalous with respect to a single one
of these measures \cite{pccp,He08,barkai2012strange}.
In our central examples FBM and diffusion on fractals have stationary
increments. In contrast, models such as CTRW or subordinated random walks
\cite{liang,fox2021aging}, both based on scale-free waiting time densities
with diverging characteristic waiting time, feature non-stationary increments.
\review{In this work we focus on processes with stationary increments, for which MSD‑based methods provide a meaningful framework for estimating anomalous‑diffusion parameters.}

Anomalous diffusion becomes more intricate in systems exhibiting
heterogeneities, which are commonly encountered in biological and
physical media. Brownian motion as well as FBM exhibit a Gaussian
distribution of displacements, a direct consequence of the central
limit theorem. However, heterogeneities can result in a non-Gaussian
distribution of displacements and variability in trajectory-to-trajectory
parameters \cite{cherstvy2019non,coppens2023anomalous,balcerek2025evaluating}. A common scenario involves trajectories that are anomalous, but
their associated parameters vary. Therefore, both $D$ and $\alpha$ are random variables and form a two-dimensional distribution
\cite{cherstvy2018,cherstvy2019non,balcerek2022fractional,maraj2025statistical}. Knowledge of such effects is crucial in data assimilation techniques aimed
at unveiling the best-fitting stochastic models given measured data.
\review{Other models that have been used in this context are FBM with random diffusivities and exponents \cite{wang2020fractional,samu,balcerek2022fractional}, random parameter Langevin equations \cite{superstatLang} or switching FBM \cite{balcerek2023,pacheco2024fractional}.}

If the diffusion parameters vary in space or time, trajectories are typically segmented into regions corresponding
to different states \cite{andi2}. While common
in experimental data analysis, the segmentation reduces the accuracy
of estimating $\alpha$ and $D$ due to the short length of individual
segments \cite{muozGil2021}. Moreover, segmentation may interfere with
existing correlations or aging properties in the system. Different
specialized methods are employed to
improve the estimation, including approaches based on machine learning
\cite{maizon2022deep,henrik1,henrik,muozGil2021,pineda2023geometric,
feng2024reliable}. Moreover,
the distribution of these subpopulations lacks a theoretical foundation, making it essential to reconstruct it as precisely as possible to better understand the heterogeneous nature of the system.
A crucial consequence is that in such setting EA-MSD and TA-MSD
become distinct, even for a process with stationary increments. The TA-MSD always measures the local parameters intrinsic to each diffusing particle, whereas the EA-MSD becomes a global average over all subpopulations and is generally not a simple power-law. Even in the simplest case with only two subpopulations with different $\alpha$, the lower exponent dominates the dynamics at short times, whereas the larger one dominates the dynamics at long times.
\review{In such heterogeneous systems, estimating the diffusion coefficient $D$ and the anomalous exponent $\alpha$ for individual trajectories becomes crucial, as ensemble averages may fail to capture the details of individual particle behaviors.}

Due to the central role that the MSD plays in the analysis of particle
trajectories, its statistical properties were given considerable scientific
attention for both normal \cite{mcCluskey,vestergaard,bullerjahn} and anomalous
\cite{kepten2013,kepten2015guide,He08,barkai2012strange} diffusion. Based on its
theoretical distribution, practical guidelines \cite{kepten2015guide} and testing
\cite{sikoraTest} were proposed. Bayesian and machine learning approaches were
also successfully applied in this area \cite{thapa2018bayesian,muozGil2021,
henrik,henrik1,andi2,munoz2020single,kowalek2019classification}. Despite this
substantial progress, the estimation of anomalous diffusion parameters remains
challenging. This problem is particularly pronounced when trajectory lengths
are short, as is usually the case in live-cell experiments. Additionally,
estimating the associated errors is non-trivial, due to the correlations
between increments within individual trajectories and the interdependence
between the estimated parameters $D$ and $\alpha$. These challenges underscore
the need for refined methodologies to reliably analyze and interpret SPT data,
ensuring accurate insights into the dynamics of complex systems.

In this work, we employ an expanded regression method to obtain improved
estimates of the diffusion coefficient $D$ and the anomalous exponent $\alpha$ in
realistic biophysical experiments. The key assumption is that each trajectory
(or the analyzed segment) is Gaussian \review{with stationary increments}, but the whole ensemble may not be,
e.g., it can involve a collection of random parameters.  We derive expressions
for the statistics of the TA-MSD in linear and log-log space. We show that for the TA-MSD, the classical linear fit in log-log space is suboptimal and a better method is proposed.

We derive formulas the for TA-MSD errors and the expected errors of the estimated
parameters for both the classical and enhanced methods, with or without experimental noise. This allows us to treat those errors as a form of  blur
distorting the original distribution of $D$ and $\alpha$, and remove it through specialised deconvolution algorithm, revealing the true structure of the particle population. We summarize the proposed statistical procedure and give examples of its application using real and synthetic data.

Implementations of the presented methods and explanations of their use are available at
\url{https://github.com/jaksle/anDiffReg}.

\section*{Methods}

\subsection*{Estimation of $\log_{10} D$ and
$\alpha$}
\label{sec:estimation}

Experimental SPT data have the form of a position time series \review{$X_i$, recorded at
times $i=0,\Delta t,2\Delta t,\ldots,n\Delta t$}. Given such data, one can obtain the corresponding TA-MSD at points $k\Delta t$
using the discrete sum \review{of squared increments}
\begin{equation}
\label{eq:tamsdDef2}
\overline{\delta^2_k}\coloneqq\frac{1}{n-k}\sum_{i=1}^{n-k}\big(X_{i+k}-X_i\big)^2.
\end{equation}
In a multi-dimensional setting, the total TA-MSD is simply the sum over all
TA-MSD components. \review{ Dealing with a discrete sum instead of the continuous
integral in Eq. \ref{eq:tamsdDef} affects the properties of the TA-MSD. We comment on the discrepancies between the definitions \ref{eq:tamsdDef} and \ref{eq:tamsdDef2} in Supplemental Material Sec.~S1.1. Further on we focus on the discrete TA‑MSD, as only the discrete form is directly obtained from experimental trajectories.}

For ergodic anomalous diffusion in a homogeneous system, we expect that, for
sufficiently long trajectories, the TA-MSD converges to the power-law form
$\overline{\delta^2_k}\approx2dD(k\Delta t)^\alpha$. Consequently, the
diffusion parameters $(D,\alpha)$ are usually obtained from a linear fit of
$\log_{10}\overline{\delta^2_k}\approx\alpha\log_{10}(k\Delta t)+\log_{10}(2dD)$ obtained by using the \emph{ordinary least squares} (OLS) method.
\review{We note that the choice of the logarithm base is inconsequential. All derivations in this paper hold for any base; for illustrative purposes, we adopt the commonly used decimal base.} Then, the vector of estimates $\widehat{\bd\beta}_\mathrm{OLS}=\boldsymbol{(}
\log_{10}(\widehat{D})+\log_{10}(2d),\widehat{\alpha}\boldsymbol{)}$ is given by the
matrix multiplication
\begin{equation}
\label{eq:olsMain}
\widehat{\bd\beta}_\mathrm{OLS}=(\bd x^\T\bd x)^{-1}\bd x^\T\bd y,
\end{equation}
where $\T$ denotes the transpose and $\bd x$ is a matrix consisting of an
all-ones column  concatenated with a column of the logarithms of time points
$\log_{10}(k\Delta t)$, while $\bd y$ is a vector of the corresponding values
$\overline{\delta^2_k}$, see Sec.~\ref{s:OLS} for details.

\begin{figure*}[ht]
\centering
\begin{tikzpicture}
\draw(0,0)node[inner sep=0]{\includegraphics[width=\textwidth]{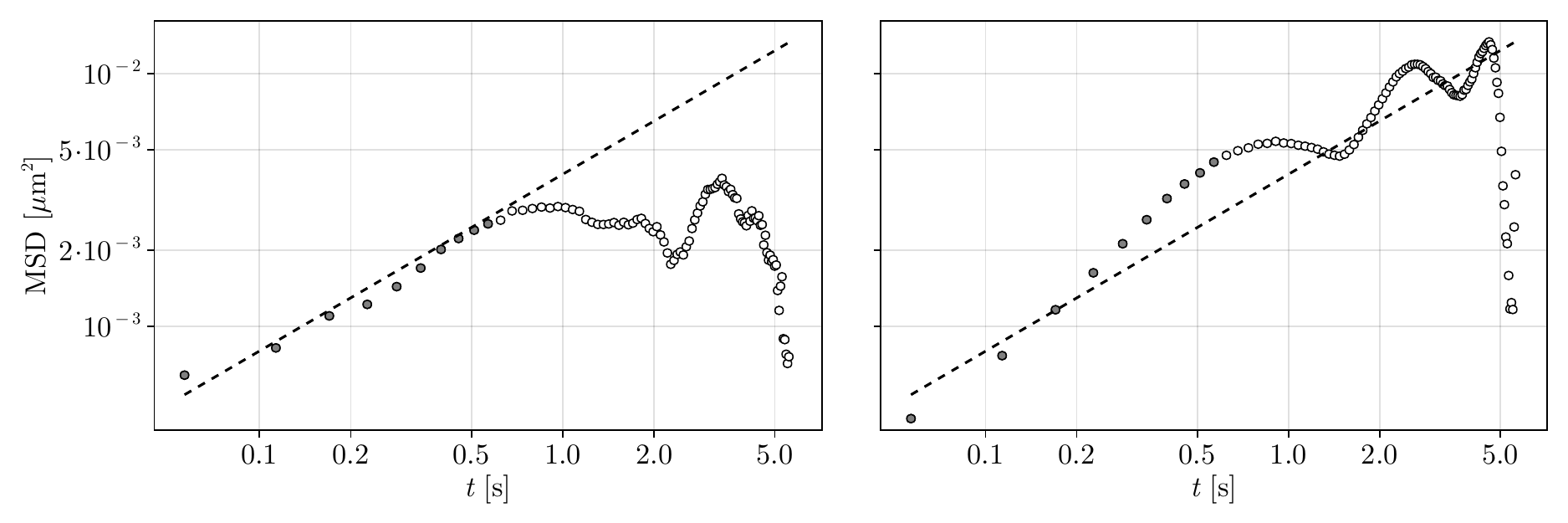}};
\draw(-7.5,2.7) node {(a)};
\draw(0.7,2.7) node {(b)};
\end{tikzpicture}
\caption{Deficiencies of standard OLS in the estimation of TA-MSD parameters. MSDs were obtained from representative simulated samples of FBM with $D=10^{-3}$, $\alpha=0.7$. Typically, only a small initial part is used for estimation, here 10\% of the data (filled circles). The dashed line shows the true MSD. (a) Trajectory with misleading long time TA-MSD. (b) Trajectory with misleading short time TA-MSD. }
\label{fig:simpleComp}
\end{figure*}

The OLS estimation is optimal when deviations from the trend are independent
and identically distributed (iid) \cite{linMod}. However, this condition is
not fulfilled for the TA-MSD. The variance of the TA-MSD errors increases with
time and these errors are correlated. Both features are clearly visible even on a
standard MSD plot, see Fig.~\ref{fig:simpleComp}. The major consequence of
breaking the OLS assumptions is that it becomes inefficient, in some cases
drastically. To avoid increasing the bias and estimation errors,
one usually discards the values of TA-MSD for times longer than $\approx 10\%$
of the total measurement time \cite{kepten2015guide}, which leaves only the more statistically reliable values, see Fig.~\ref{fig:simpleComp}(a). At the same time, this approach neglects a large part of the available information. For some trajectories, using long time values would correct the misleading short-time trend, see Fig.~\ref{fig:simpleComp}(b).

In order to account for the dependence of the TA-MSD values at different time
points, we employ the \textit{generalized least squares} (GLS) method
\cite{linMod, econometrics}, designed for linear regression problems in the presence of
non-iid deviations. With the same notation as above, the GLS estimate is given
by 
\begin{equation}\label{eq:glsMain}
\widehat{\bd\beta}_\mathrm{GLS}=(\bd x^\T\Sigma^{-1}\bd x)^{-1}\bd x^\T\Sigma^{
-1}\bd y,
\end{equation}
where $\Sigma$ is the log TA-MSD error covariance matrix, see  Sec.~S1.4 for details. The idea behind the GLS method is the following: given $\Sigma$ we
linearly transform the data to $\sqrt{\Sigma^{-1}}\bd x$ and $\sqrt{\Sigma^{-1}}
\bd y$, so that the new coordinates have iid deviations, and thus an efficient
fit can be obtained. For anomalous diffusion, this transformation can be considered to be a form of fractional derivative \cite{meerschaert2011stochastic} as it removes power-law memory from the log TA-MSD time series.

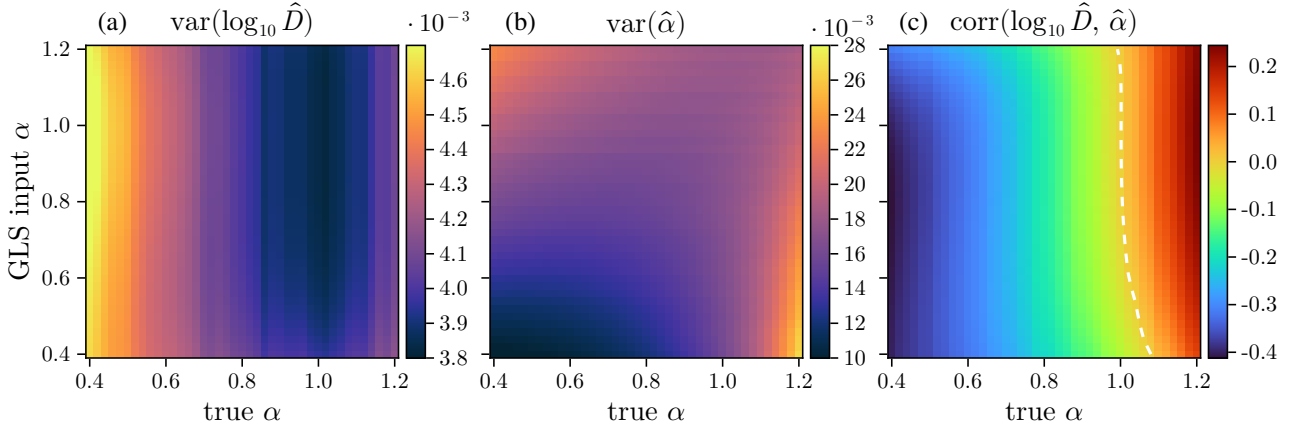
\begin{figure} \centering
	\begin{tikzpicture}
		\draw(0,0) node[inner sep=0]{\includegraphics[width=1\textwidth]{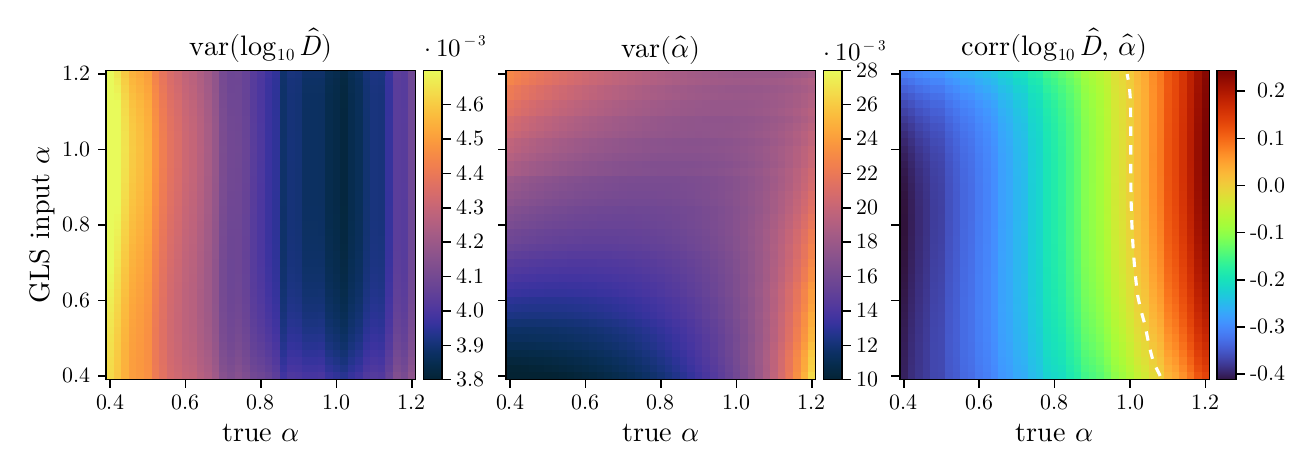}};
		\draw(-7.0,2.5) node {(a)};
		\draw(-1.6,2.5) node {(b)};
		\draw(3.6,2.5) node {(c)};
	\end{tikzpicture}
	\caption{\review{Monte Carlo–based sensitivity analysis of the feasible GLS with respect to the input $\alpha$ values.  (a) variance of $\log\widehat{D}$, (b) variance of $\widehat{\alpha}$, (c) correlation of  $\log\widehat{D}$ and $\widehat{\alpha}$ estimates; the white dashed line denotes zero correlation.
    The data were simulated from FBM with $n=100$, $D=1/2$,
		and $\Delta t=1$. Sample size was $10^5$ trajectories. }
        }
	\label{fig:inputH}
\end{figure}

The GLS method requires the knowledge of the matrix $\Sigma$. Crucially,
assuming that the data correspond to power-law correlated anomalous diffusion
is sufficient to adequately describe its form, Eq.~S4. Matrix $\Sigma$ depends
only on the anomalous exponent $\alpha$. This exponent is usually unknown, but a preliminary estimate can be used, yielding an estimate $\widehat{
\Sigma}$. The resulting GLS is a type of the so-called \emph{feasible GLS} approach \cite{econometrics}. \review{In Fig. \ref{fig:inputH} we present a Monte Carlo based sensitivity analysis of the feasible GLS method with respect to the input $\alpha$. The obtained values of the variances of the estimators confirm robustness of the method in the experimentally relevant range
$0.4 \leq \alpha \leq 1.2$. } 

\begin{figure}\centering
\begin{tikzpicture}
\draw(0,0) node[inner sep=0]{\includegraphics[width=1\textwidth]{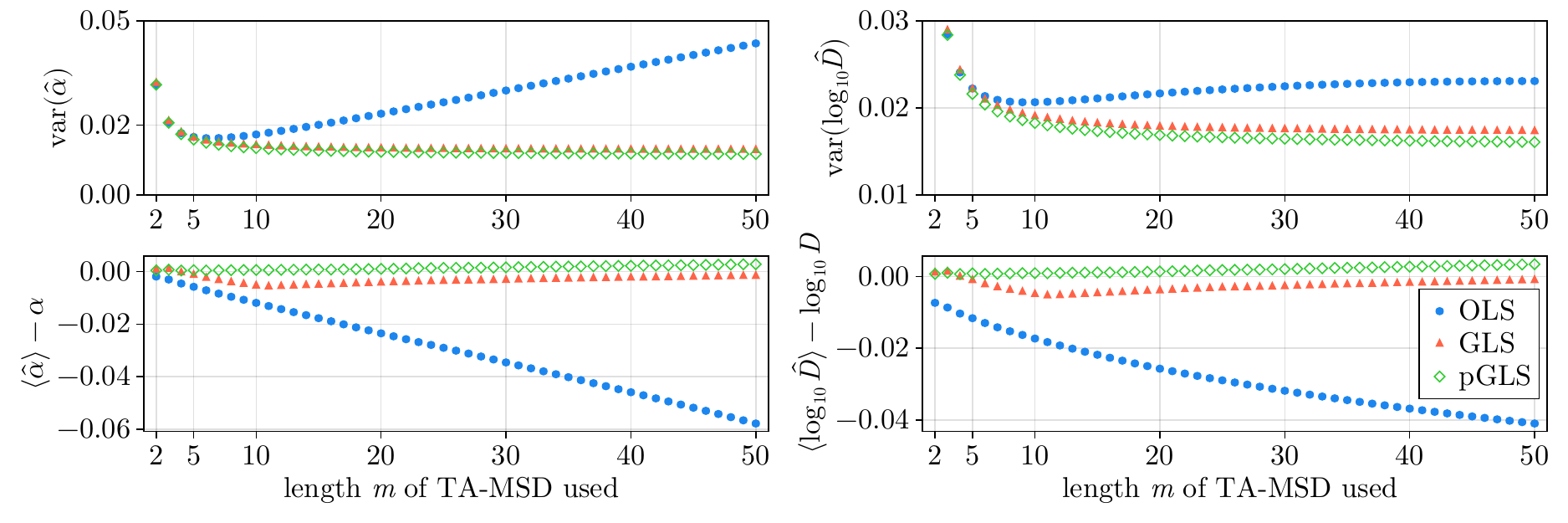}};
\draw(-7.8,3.1)node {(a)};
\draw(0.9,3.1)node {(b)};
\draw(-7.8,0.3)node {(c)};
\draw(0.9,0.3)node {(d)};
\end{tikzpicture}

\caption{Numerical comparison of the estimation quality for the OLS, \review{(feasible) GLS with input $\alpha$ estimated by OLS with a window length of $m$, if $m\ge$10, and a window length of 10 otherwise, and perfect GLS (pGLS) with
 true (non-estimated) $\alpha$ used}. The data involve FBM trajectories of length 100, $\Delta t = 0.0567$ and $\alpha = 0.5$. (a) Variance of $\widehat\alpha$. (b) Variance of $\log_{10}\widehat D$. (c) Bias of $\widehat\alpha$
. (d) Bias of $\log_{10}\widehat D$. Sample size was $10^5$ trajectories. }
\label{fig:lenComp}
\end{figure}
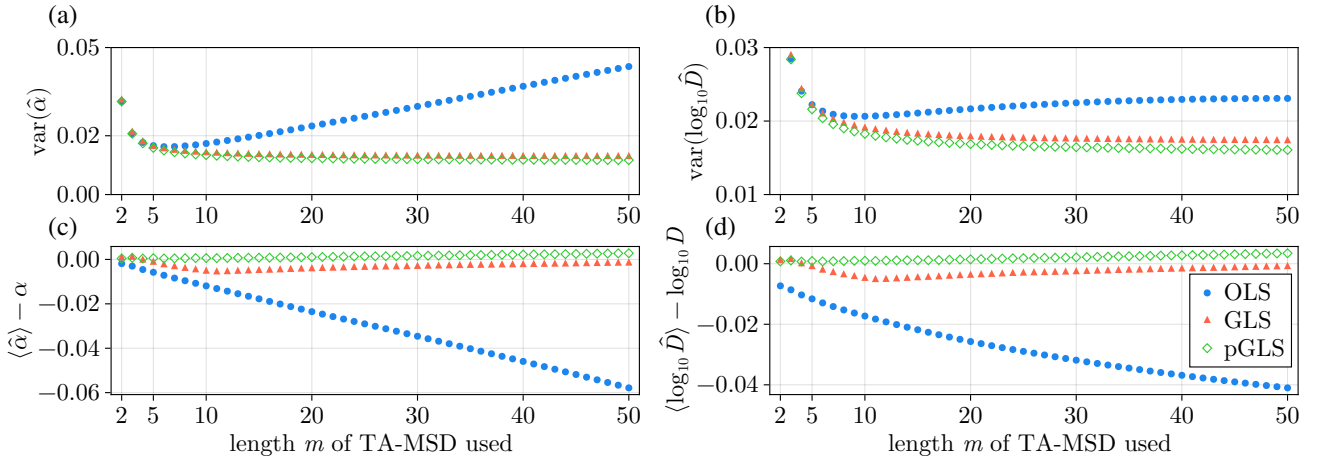

\review{In Fig. \ref{fig:lenComp}, we compare the performance of the OLS, feasible GLS and GLS with perfect $\alpha$ (pGLS) methods for different TA-MSD window lengths $m$. The input value of $\alpha$ for the feasible GLS is obtained from an OLS estimate computed over a 10‑point window. The feasible GLS performs nearly as well as the pGLS, which is the optimal possible regression estimator. 
The advantage of the GLS approach becomes especially pronounced for larger window sizes; this method can and should use all of the TA-MSD values. 
For OLS the optimal number of points depends on both the model and trajectory length. GLS always uses all information effectively, and therefore outperforms the best-case OLS. The remaining small bias of the feasible GLS is inherited from the initial window-10 OLS and gets corrected as the GLS window increases. The minor bias of the perfect GLS is due to the covariance approximation and can be removed using the exact formula, but at much higher computational cost. }

Using  the fact that \ref{eq:olsMain} or \ref{eq:glsMain} are linear equations and we know $\widehat \Sigma$, the amplitudes of the estimation errors for both diffusion parameters can be established for each analyzed trajectory separately (see Eqs.~\ref{eq:olscov} and \ref{eq:glscov}). These estimates are crucial for further analysis of
the system, such as identifying subpopulations. The derived theoretical
formulas predict strong correlations between estimates of $\alpha$ and $\log D$ for both the OLS and GLS
(see Eq.~\ref{eq:OLSrough} and the corresponding discussion), though in the case of the GLS they are less extreme. It is a purely
statistical effect that needs to be taken into account when trying to decipher the physical
structure of the ensemble. \review{In Fig. \ref{fig:inputH}c we illustrate the correlation for simulated FBM trajectories. }

\subsection*{Reconstructing
the joint distribution of $D$ and $\alpha$}
\label{s:recon}

The methods described in the previous section allow one to estimate $\log D$
from individual trajectories. However, recovering the diffusivity $D$ requires
some additional attention due to the non-linear intrinsic relation between the
diffusivity and the estimated value $\log D$, present for both OLS and GLS
methods. The resulting bias in the estimation of $D$ can be calculated using the
estimated error variances obtained from our procedure, see Sec.~S3.1.
Using these values, we can mitigate the bias by appropriately correcting the
diffusivity through $\widehat{D}_\mathrm{cor}=\widehat{D}\mathrm{e}^{-w/2}$,
where $w=(\ln(b)e_1)^2$, $b$ is the chosen base of the logarithm used, $\ln$
is the natural logarithm, and $e_1$ is the standard error of the estimated
value $\log_b\widehat D$. Here, for illustration of the methods we use the
decimal logarithm with $b=10$; \review{results obtained with other bases would be scaled accordingly.} 

Using the estimators of $\log D$ and $\alpha$ for individual trajectories one obtains a sample of estimates ($\log_{10}\widehat{D},\widehat{\alpha})$. This sample does not have the same distribution as the original ($\log_{10}D, \alpha$). This is because the statistical errors distort the real population structure and their influence can be considered a form of blur. Because the errors are correlated, they introduce an artificial correlation between the estimates. Hence, it is important to include the information about the error
distribution when analysing joint distribution of $(\log D, \alpha)$ estimated from the TA-MSD data.

\review{For fixed true $(D,\alpha)$ the estimators $(\log_{10} \widehat D,\widehat \alpha)$ for both the OLS and GLS have Gaussian distribution when calculated from long trajectories (the mathematical argument is shown in Sec.~\ref{s:gauss}).} In practical settings, a length of 50 points is enough for this approximation to work well. Because we also established the error bias and covariance, we know all the parameters of the blur and we can remove its influence. Note that it is more
convenient to work with the logarithm of the diffusivity at this stage because
it renders the distribution spread lower and the error distribution becomes
independent of the value of the diffusivity, see Sec.~\ref{s:reconstruct} for
details.

The real distribution can be recovered by performing a deconvolution. In simple
cases, such as Gaussian distributions, analytical formulas can be derived, see
Sec.~\ref{s:pDeconv}. For more complicated populations, such as our experimental
data considered below, only numerical methods are feasible. To this end, we
propose to employ the Richardson--Lucy deconvolution, a simple and effective
iterative algorithm widely used in image processing to restore images blurred by
a known point spread function \cite{richardson1972bayesian,lucy1974iterative}.
The entire statistical procedure is summarized as an itemized list in
 Supplemental Material Sec.~\ref{s:statSum}.

\section*{Results}

\subsection*{Numerical simulations}
\label{s:sim}

\review{To test our statistical procedure, we simulated 3 sets of trajectories designed to resemble typical SPT data.} The first example are length 100 FBM trajectories. FBM is a process found in a vast array of SPT experiments and deeply studied theoretically \cite{christine,weber,weakErgJeon,jae_prl,lene1,vilk2022unravelling,burnecki,magdziarz,
sabri2020elucidating}. It is an ergodic process \cite{deng,pccp} driven by
long-ranged, power-law correlated Gaussian noise \cite{mandelbrot1,kolmogorov}.
However, our method can be used for different models as well.
This example is supposed to replicate the essential aspects of the data analysed in the next section, so we first simulated two-dimensional FBM trajectories with
parameters (diffusivity, anomalous exponent, time range, number of measured
data points) in a similar range as those measured in typical intracellular
environments \cite{sabri2020elucidating}. Our results can be directly applied
to the analysis of experimental data, as shown in the next Section. We
included three particle subpopulations: superdiffusive with $\alpha=1.4$, subdiffusive
with $\alpha=0.5$, and immobile. The latter was modeled as a simple
iid Gaussian time series. For each case we simulated 10,000 trajectories.

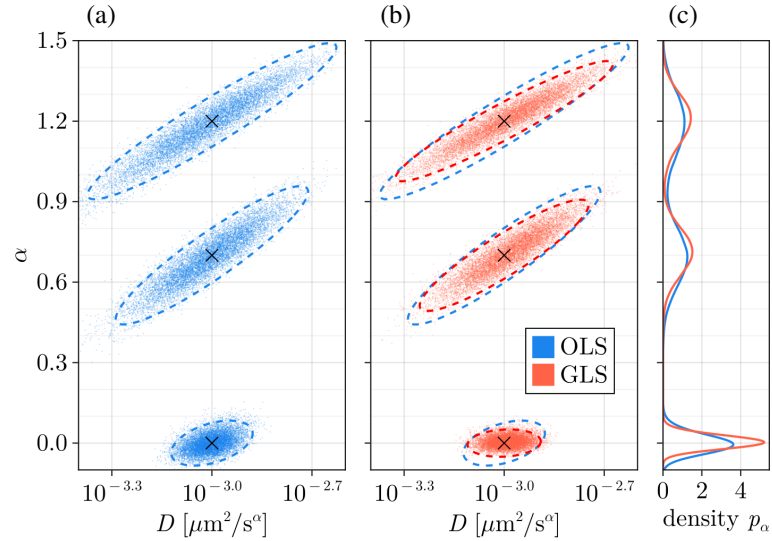
\begin{figure}
\centering
\begin{tikzpicture}
\draw(0,0) node[inner sep=0]{\includegraphics[width=0.6\textwidth]{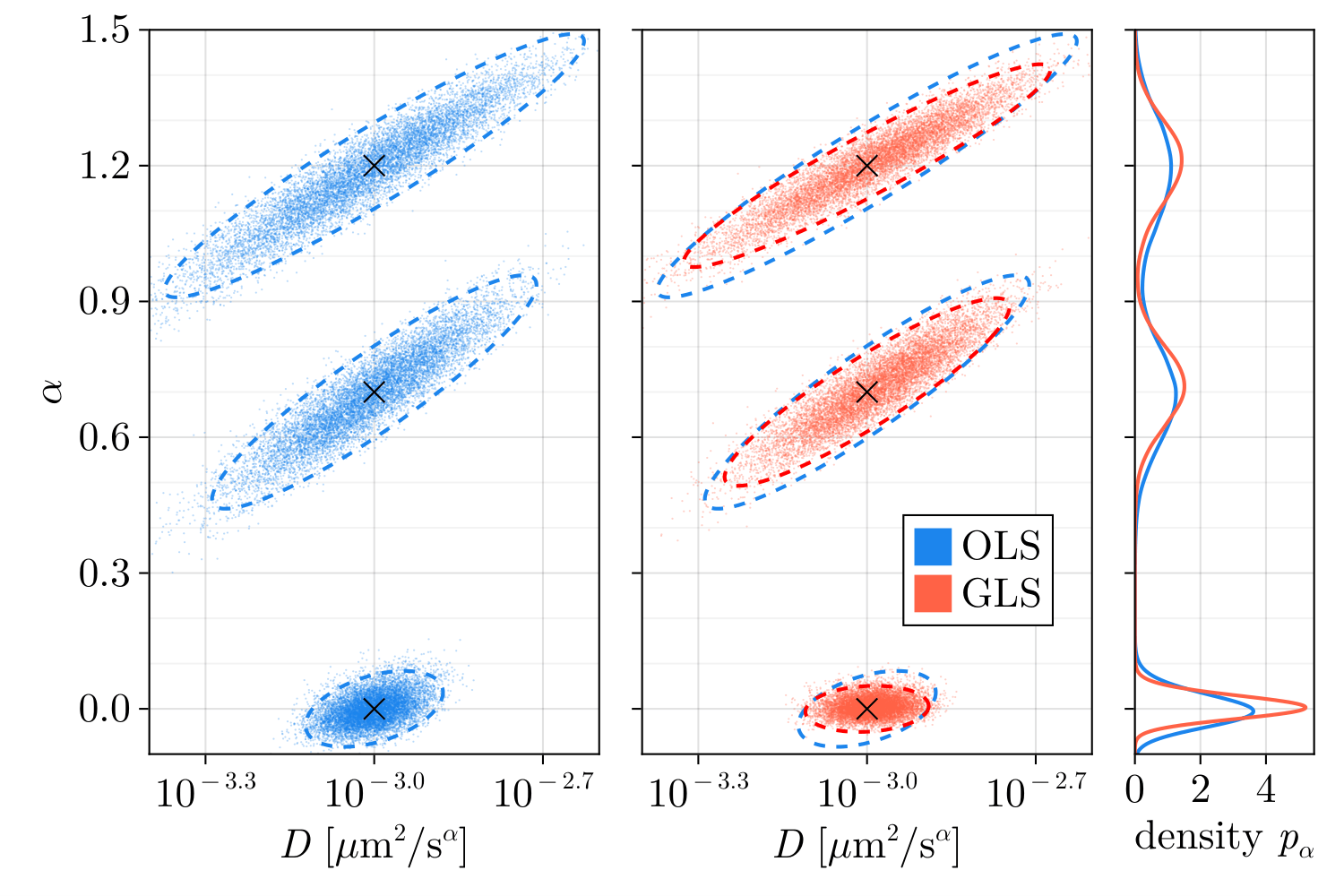}};
\draw(-3.8,3.6)node {(a)};
\draw(0.1,3.6)node {(b)};
\draw(3.9,3.6)node {(c)};
\end{tikzpicture}

\caption{Joint estimation of diffusivity $D$ and anomalous diffusion exponent $\alpha$
for two subpopulations of simulated FBM trajectories with $\alpha=0.7$ and $\alpha=1.2$ (marked by crosses),
and a subpopulation of immobile tracers, with $\Delta t=0.0567$, length 100 trajectories. (a) Estimation using OLS. Dashed
ellipses denote the predicted $95\%$ confidence regions. (b) Estimation using GLS. Confidence regions are shown for both OLS (in blue) and GLS (in red), for comparison. (c) Estimated PDF of the measured $\alpha$ values. Sample size was $10^5$ trajectories; for clarity only $10^4$ points are shown. }
\label{fig:ellipses}
\end{figure}

The estimation results using both the classical linear fit in log space (the
OLS) and the proposed GLS method for 100-point trajectories are shown in Fig.~\ref{fig:ellipses}. For the OLS estimation we used only the 10 first TA-MSD values, while GLS was
applied to all of them. The theoretical $95\%$ confidence areas, calculated
from Eqs.~\ref{eq:olscov} and \ref{eq:glscov} are also depicted in
Fig.~\ref{fig:ellipses}. Although the formulas for the theoretical estimates of the
confidence areas are approximates supposed to work for sufficiently long
trajectories, it is clear that they match the results of the simulations
very well. In each of the considered cases, the GLS errors are substantially
smaller than the OLS counterparts for both the diffusion coefficient and the
anomalous exponent $\alpha$. These observations provide evidence for
the improvement using the GLS method for TA-MSD analysis instead of OLS.
Fig. \ref{fig:ellipses} shows the observed
scatter of the estimates of $D$ and $\alpha$. This scatter is concentrated
on a line in the $D,\alpha$ phase space, which means that the estimation
introduces strong artificial correlations of these parameters within each
subpopulation. The sign of the correlation depends on the value of the
time step chosen as $\Delta t=0.0567$, see Eq.~\ref{eq:OLSrough}.

\begin{figure}
	\centering
	\begin{tikzpicture}
		\draw(0,0) node[inner sep=0]{\includegraphics[width=1.\textwidth]{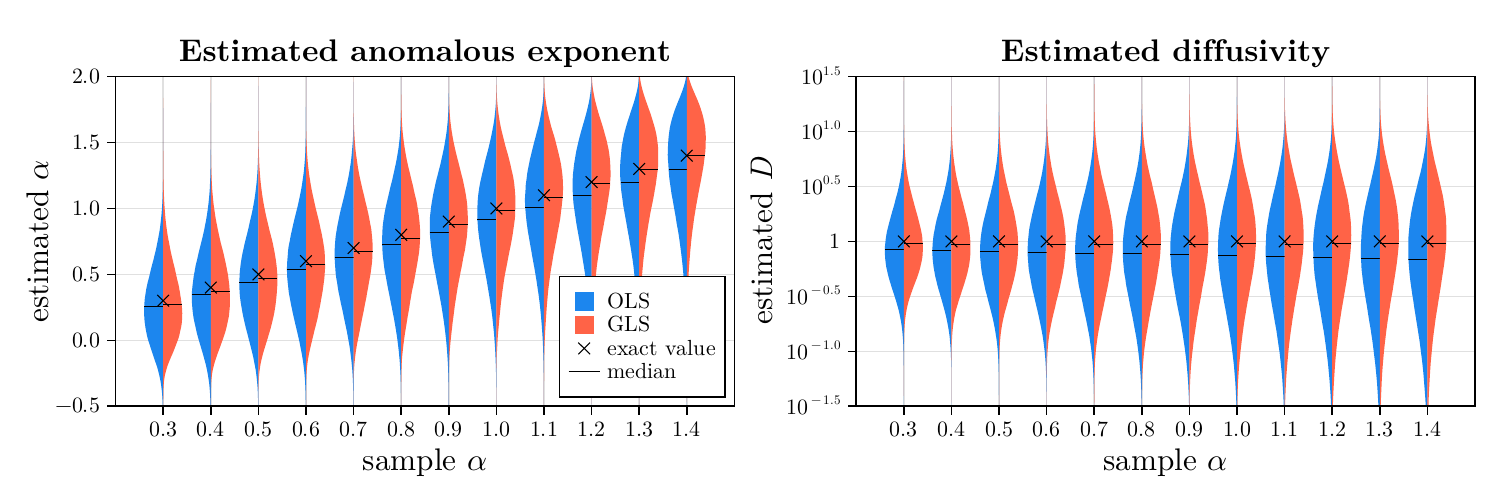}};
		\draw(-7.2,2.3) node {(a)};
		\draw(1.5,2.3) node {(b)};
	\end{tikzpicture}
	
	\caption{Violin plots of OLS versus GLS method for extremally short 10-point trajectories of FBM. (a) Plots for anomalous exponent $\alpha$,  (b) for diffusivity $D$. The OLS was fitted using 5 TA-MSD points, the GLS used all 9 TA-MSD points available. Sample size was $10^5$ trajectories.}
	\label{fig:violin}
\end{figure}

\review{
As the second example, we tested the feasible GLS on ultra‑short FBM trajectories, which may arise, e.g., in time‑ and space‑dependent diffusion analyses where longer trajectories are segmented into shorter pieces. We consider 10‑point simulated FBM trajectories, for which the TA‑MSD contains only nine values. In Fig. \ref{fig:violin} we compare the 5-point OLS estimate to the feasible GLS which uses all 9 available points. Two effects are visible: GLS substantially reduces the bias and yields thinner lower tails, in particular yielding fewer unphysical cases with $\alpha \le 0$.  Note also that an OLS bias of this magnitude implies that diffusivity estimates may be skewed by up to $40\%$ for superdiffusive motion, whereas for GLS the deviation is at most about $6\%$. Additional details are presented in the Supplementary Material, Sec.~\ref{s:short}.
}

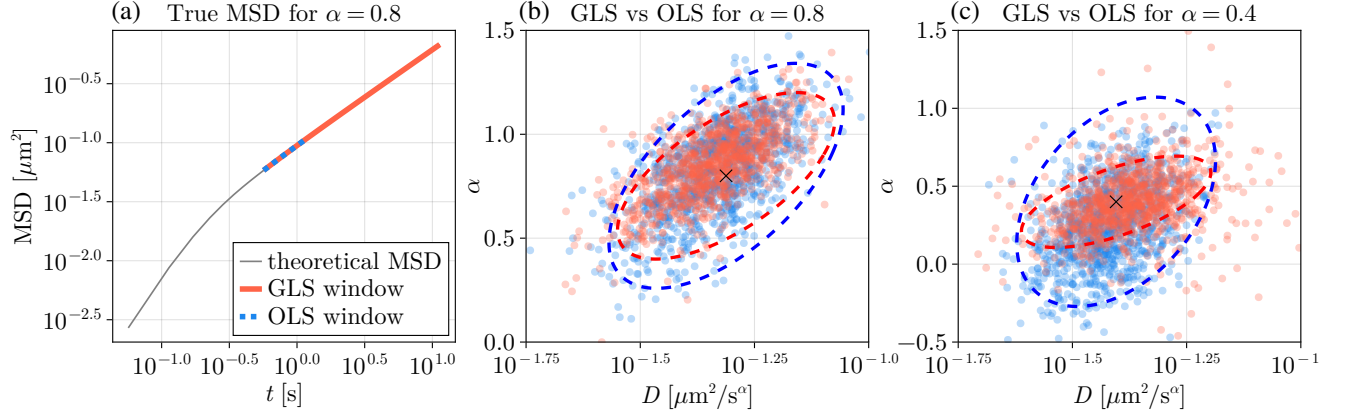
\begin{figure}[ht]
\centering
\begin{tikzpicture}
\draw(0,0) node[inner sep=0]{\includegraphics[width=1\textwidth]{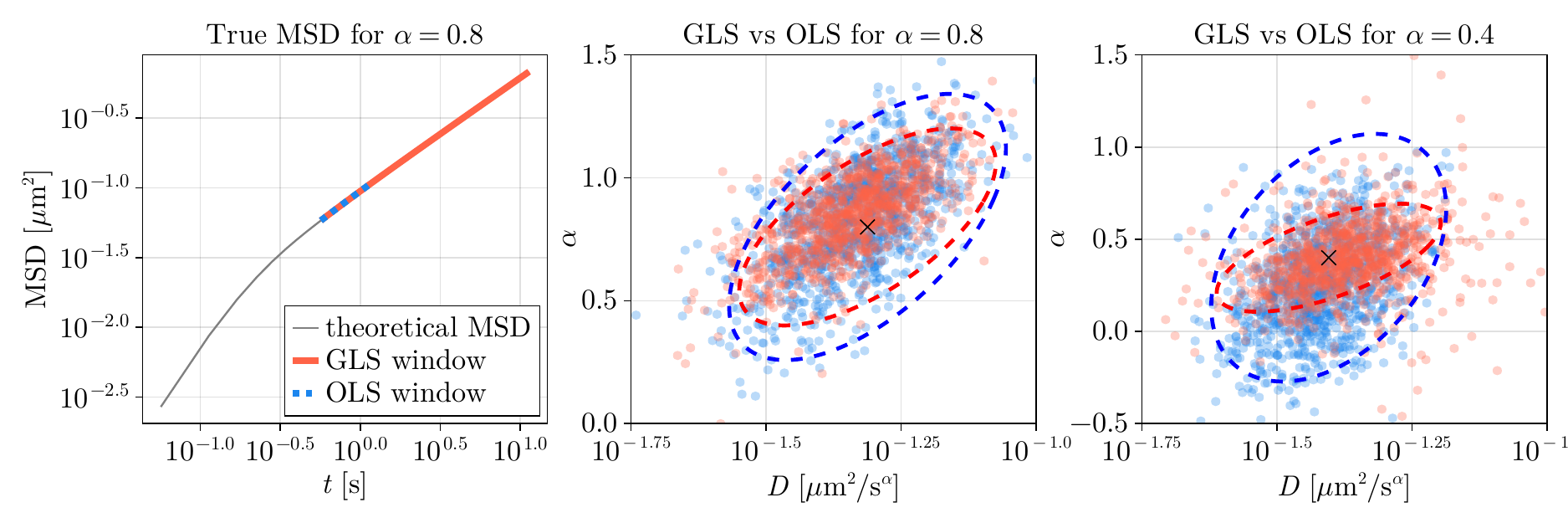}};
\draw(-7,2.55) node {(a)};
\draw(-1.6,2.55) node {(b)};
\draw(4.1,2.55) node {(c)};
\end{tikzpicture}
\caption{\review{ (a) OLS and GLS estimation windows marked on the theoretical MSD of GLE solution. (b) and (c) Comparison of OLS (blue points) and GLS (red points) estimates for GLE with $\zeta=20$ and $\alpha = 08$, $\alpha = 04$ respectively. Black crosses are  true values. Dashed
ellipses denote the $95\%$ confidence regions predicted for OLS and perfect GLS. Sample size was $10^4$ trajectories. } }
\label{fig:GLE}
\end{figure}

\review{As the last, third example, we check how the procedure works for the solutions of GLE with power-law kernel (also called "fractional Langevin equation" in the literature) which is  another key model of diffusion in viscoelastic media and can be derived from fundamental principles of statistical physics \cite{kouPRL, kouAAS,lutz}. At long timescales this model becomes indistinguishable from FBM, but at short times the motion is ballistic. For a demonstration we chose the friction constant to be $\zeta = 20$ for which at sampling rate $\Delta t=0.0567$ the motion is visibly distinct from FBM, yet there is still a range in which the diffusion parameters can be fitted without substantial bias. Such conditions are a challenge to regression-based methods, because one cannot use initial TA-MSD values which are affected by the ballistic regime, yet what remains has significantly worse statistical reliability. In Fig.~\ref{fig:GLE} we demonstrate that this situation results in GLS gains becoming more pronounced, with variance of $\alpha$ estimates decreasing by 40\% in panel (b) and 15\% in panel (c). In panel (c) the OLS returns 20\% of physically impossible negative values of $\alpha$ whereas for the GLS it is only 3\%. The dashed ellipses are theoretically predicted confidence regions for OLS and perfect GLS. For the hardest case in (c) feasible GLS gained 50\% of $\widehat\alpha$ variance compared to pGLS which was due to the very poor quality of the preliminary estimates taken from OLS. It was still 15\% better than the OLS, but it shows that if a reasonable range of the estimated parameters is proposed this estimator can be further improved. Additional information about the model and simulations are available in the Supplement, Sec.~\ref{s:GLE}.
}

Additionally, in Supplemental Material Sec.~\ref{s:noise} we also provide theoretical and numerical analysis
of the estimators' behaviors in the presence of experimental
noise. It shows that when it is present the GLS provides even more significant reduction to bias and variance.

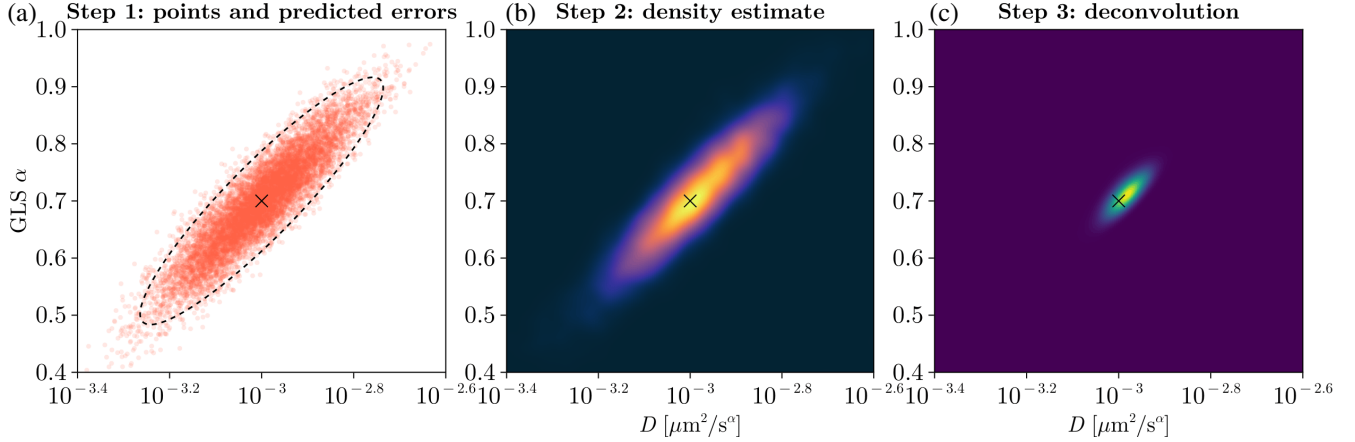
\begin{figure}[ht]
\centering
\begin{tikzpicture}
\draw(0,0) node[inner sep=0]{\includegraphics[width=1\textwidth]{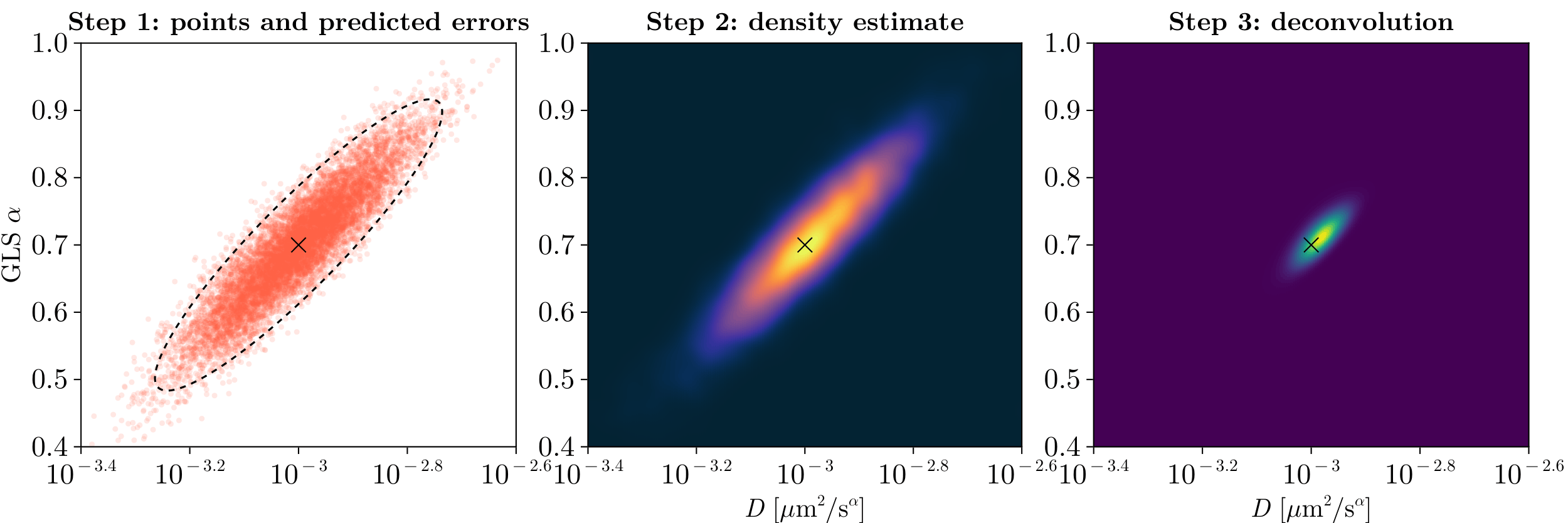}};
\draw(-8.6,2.65) node {(a)};
\draw(-2.0,2.65) node {(b)};
\draw(3.6,2.65) node {(c)};
\end{tikzpicture}
\caption{Deconvolution applied to simulated FBM trajectories of length 100
with $\Delta t=0.0567$, $D=10^{-3}$ and $\alpha=0.7$ (marked by cross). The PDF was estimated with the kernel density method.(a) Scatter plots of GLS estimated values of $\alpha$ and $D$. Dashed line are the predicted 95\% confidence regions. (b) Joint PDFs of the estimates of $\alpha$ and $D$. (c)  Joint PDFs after deconvolution with the Richardson-Lucy algorithm. Sample size was $10^4$ trajectories. }
\label{fig:decEx}
\end{figure}

As the last part of numerical tests, we applied the deconvolution procedure 
to the estimates obtained in the first example, that is,
values $(\log_{10}\widehat{D},\widehat{\alpha})$ estimated from from 100-point 2D FBM trajectories. Fig.~\ref{fig:decEx} shows the joint
probability density function (PDF) of the parameters $D$ and $\alpha$ estimated from a single population of FBM trajectories with $\alpha=
0.7$. We used the formula for the error covariance (Eq.~\ref{eq:sigma}) to remove the spread related
to the estimation errors and not to the original ensemble. The resulting PDF is
also plotted in Fig.~\ref{fig:decEx}. The application of the
deconvolution significantly improves the estimation of the original joint
PDF, reducing the artificial variance several times.

\subsection*{Experimental data: quantum dots in the cytoplasm of mammalian
cells}
\label{s:data}

\begin{figure}[ht]
\centering
\begin{tikzpicture}
\draw(0,0) node[inner sep=0]{\includegraphics[width=1\textwidth]{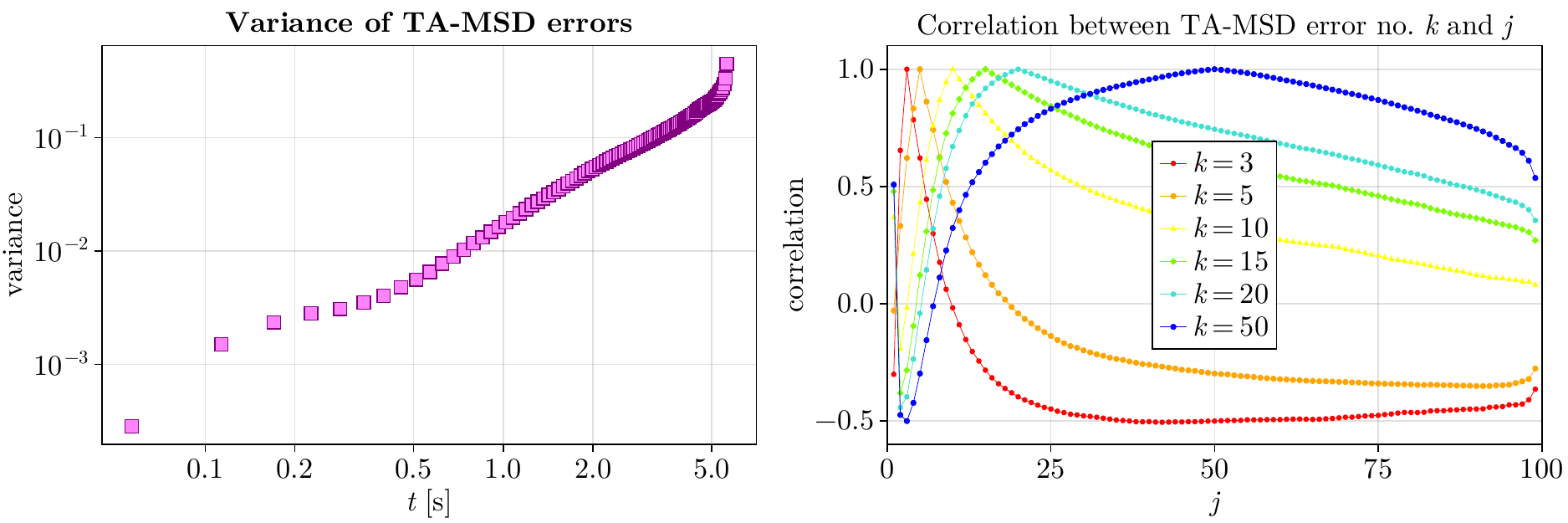}};
\draw(-8,2.7)node {(a)};
\draw(0.8,2.7)node {(b)};
\end{tikzpicture}

\caption{The conditions for efficient classical regression are broken for anomalous diffusion experimental data. The averages were calculated over the entire 4,844 trajectory dataset. (a) Average variance of log TA-MSD. The error variance increases by more than two orders of magnitude as the lag time $t$ increases. (b) Average correlation of log TA-MSD errors. The correlation depends on labels of $\overline{\delta^2_j}$ and $\overline{\delta^2_k}$.}
\label{fig:err}
\end{figure}

\review{We demonstrate the application of our method in experimental data analysis by characterizing two-dimensional SPT trajectories of semiconductor nanocrystals}  (i.e., quantum dots) that were introduced into the cytoplasm of live HeLa (human cervical cancer) cells. The incorporation of the quantum dots was achieved by bead loading and imaged in a custom-built fluorescent microscope as previously described \cite{sabri2020elucidating,janczura2021identifying}. A
total of 4,844 trajectories, each consisting of 100 points, were analyzed. We used a two-dimensional TA-MSD, so the parameters were fitted to the formula $4Dt^\alpha$ and all the estimation errors were twice as high as compared to
a one-dimensional trajectory. First, we checked the OLS assumptions for these data. The variance of the errors, their correlations and the spread of the TA-MSD curves are plotted in Fig.~\ref{fig:err}. The deviations from
the proper OLS setting are drastic, meaning that one should be very careful
when using this method.

\begin{figure*}[ht!]
\centering
\begin{tikzpicture}
\draw(0,0) node[inner sep=0]{\includegraphics[width=1\textwidth]{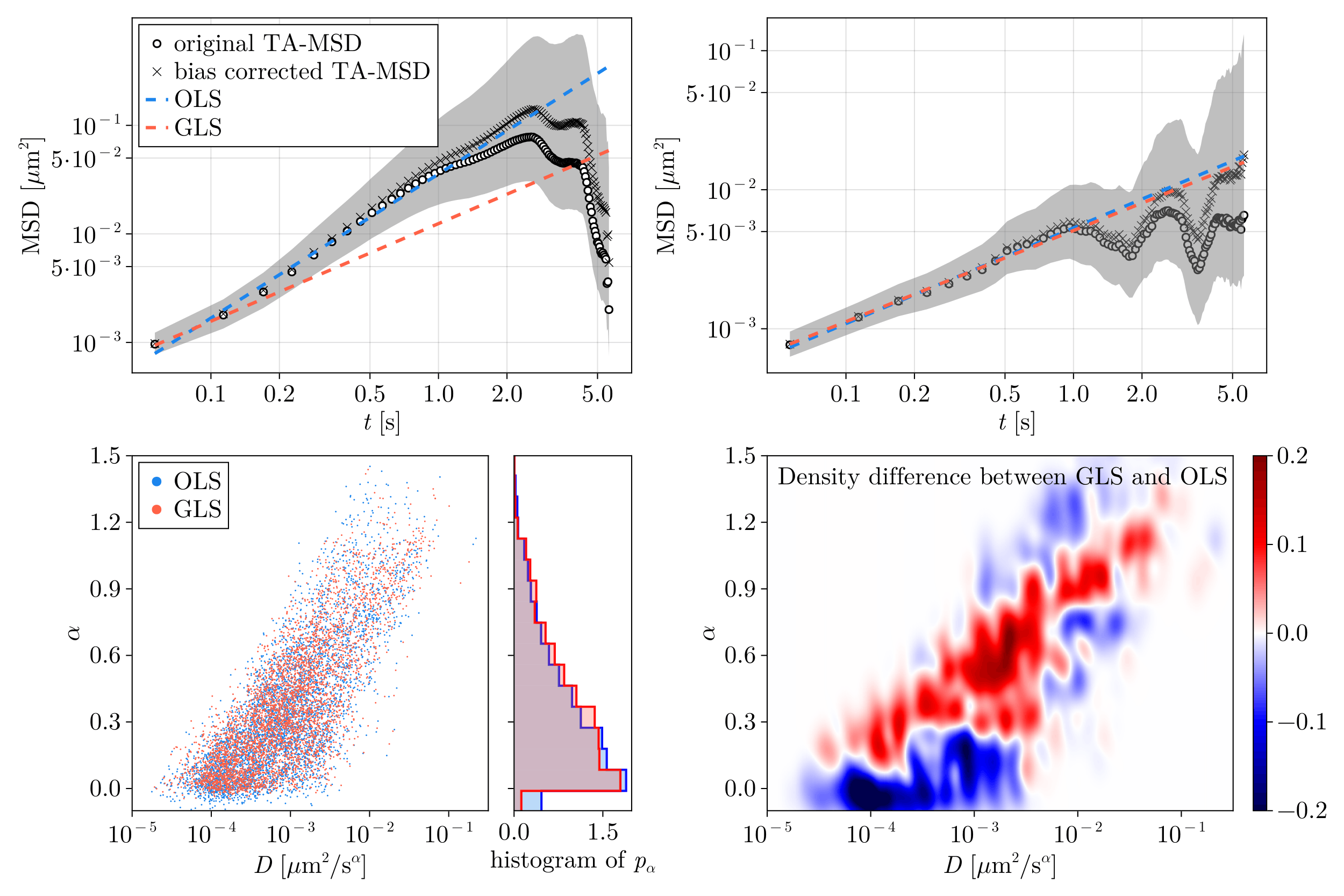}};
\draw(-7.5,5.5) node {(a)};
\draw(0.5, 5.5) node {(b)};
\draw(-7.5,0.5) node {(c)};
\draw(0.5,0.5) node {(d)};
\end{tikzpicture}
\caption{Analysis of quantum dot trajectories in the cytoplasm of live HeLa
cells. (a,b) TA-MSDs of two representative experimental trajectories
together with their OLS and GLS fits. Grey areas show $\pm$TA-MSD standard
deviations (Eq.~S2). (c) Scatter plot of the estimated
$D$ and $\alpha$ values together with the marginal histogram of $\alpha$
for the entire experimental sample. (d) Difference between the densities of
the estimates obtained using the GLS and OLS methods. The densities were
assessed by a kernel estimator. The total difference amounts to 16\% of the
probability mass.}
\label{fig:data}
\end{figure*}

Next, we applied the GLS estimation. A linear OLS fit based on the first 10 TA-MSD points was employed to obtain the initial guess of $\alpha$ and thus the error covariance matrix $\Sigma$. For trajectories with estimated $\alpha$
below a threshold of 0.1, the value of 0.1 was used as the initial guess. \review{This threshold is introduced for numerical stability, since the FBM covariance matrix becomes singular as $\alpha\to 0$. Moreover for extremely subdiffusive trajectories the OLS becomes unreliable as it often returns negative $\hat\alpha$ values, so for such cases fixed initial $\alpha$ becomes a more effective choice.} Subsequently, we employed all available 99 TA-MSD values and obtained final GLS estimates. The results of the procedure are shown in Fig.~\ref{fig:data}. 
The violations of the assumptions of the OLS method are
visible on the TA-MSD in Figs.~\ref{fig:data}(a) and (b): the smooth meandering of
the TA-MSD with increasing windings means its values are correlated and have
increasing variance. Consequently, the GLS estimation yields an improvement,
particularly when the initial part of the TA-MSD is not representative for its
overall trend, such as in Fig.~\ref{fig:data}(a). For some trajectories, the
estimated anomalous exponents using both methods OLS and GLS appear consistent,
such as in Fig.~\ref{fig:data}(b), but such cases are not representative. The
corresponding OLS and GLS estimates of $D$ and $\alpha$ for the two trajectories
presented in Figs.~\ref{fig:data}(a) and (b) are summarized in Table~\ref{tab:fitRes}.

\begin{table}
\centering
\begin{tabular}{|c|cccccc|}
\hline
Figure & $\log_{10}\widehat D_\mathrm{OLS}$& $\log_{10}\widehat D_\mathrm{GLS}$ & $\widehat D_\mathrm{OLS}$ & $\widehat D_\mathrm{GLS}$  & $\widehat \alpha_\mathrm{OLS}$ &$\widehat \alpha_\mathrm{GLS}$ \\ \hline
\ref{fig:data}a) &   -1.74$\pm$0.14&   -1.51$\pm$0.13& $ 0.17\pm0.06$  & $0.30\pm0.09$ & 0.83$\pm$0.12&    1.08$\pm$0.091\\
\ref{fig:data}b) &   -2.88$\pm$0.11&   -2.90$\pm$0.10& $0.013\pm0.003$ & $0.012\pm0.003$ & 0.69$\pm$0.10&      0.66$\pm$0.083\\\hline
\end{tabular}
\caption{Estimates corresponding to the two trajectories with MSDs shown in
Figs.~\ref{fig:data}(a) and \ref{fig:data}(b). Diffusivities have units $[\unit{\um}^2/
\mathrm{s}^{\alpha}]$. The errors are standard deviations calculated using
relations S8 and S13. Values of $\widehat{D}$ are
bias-corrected according to Tab.~S2.}
\label{tab:fitRes}
\end{table}

Overall, employing the GLS estimation changes the distribution of diffusivities
and anomalous diffusion exponents (Figs.~\ref{fig:data}(c) and (d)). When using
the GLS method, the distributions of $D$ and $\alpha$ exhibit less spread and
are concentrated within smaller ranges. Importantly, there are much less
trajectories with negative-valued outcomes for $\alpha$ which are likely not
physical and emerge only due to statistical errors. This effect allows for a
better identification of trajectories with a truly small $\alpha$, which are
likely a subpopulation of immobile or partially immobile particles. The second
important observation is that when GLS is used, superdiffusion is less
pronounced, i.e., there are fewer trajectories with $\alpha>1$, and those
remaining have substantially smaller $\alpha$ exponents than the results
using the OLS method.

\begin{figure*}
\centering
\begin{tikzpicture}
\draw(0,0) node[inner sep=0]{\includegraphics[width=0.75\textwidth]{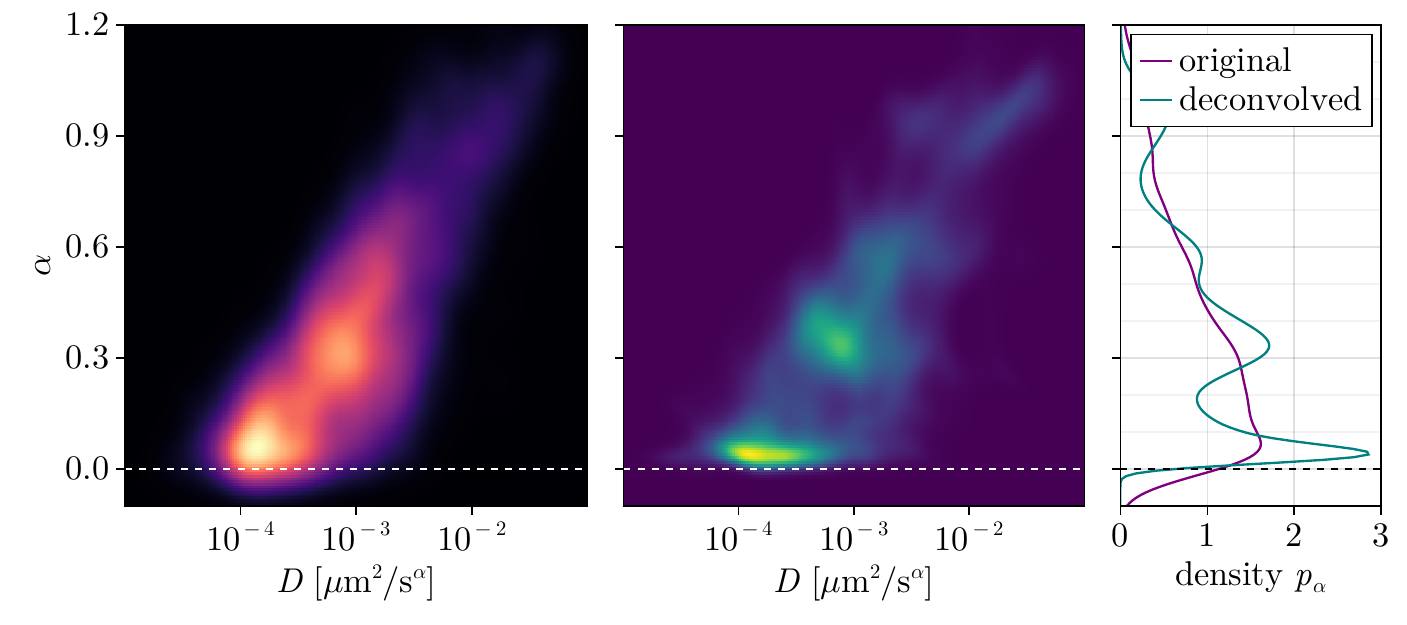}};
\draw(-5.2,2.9)node {(a)};
\draw(-0.5,2.9)node {(b)};
\draw(4.2,2.9)node {(c)};
\end{tikzpicture}
\caption{Reconstruction of the joint PDF of $D$ and $\alpha$ for the quantum dot
data. (a) PDF of the values estimated using the GLS method. The PDF was
calculated using a kernel density. (b) PDF after deconvolution using a
Richardson--Lucy iterative algorithm with local $\alpha$; 100 steps were used after which the
result stabilized. (c) Marginal densities of $\alpha$ estimated with and
without deconvolution.}
\label{fig:dataDecon}
\end{figure*}

Finally, we applied the iterative deconvolution algorithm to the joint
distribution of the diffusion parameters obtained from the GLS. In Fig.~\ref{fig:dataDecon}(a),
we show the joint PDF directly obtained from the GLS estimation. One of the first noticeable features is the appearance of non-physical negative $\alpha$ values on the low side of a cluster of trajectories with very small $\alpha$ and $D$. As mentioned above, this group of trajectories likely corresponds to trapped or immobilized particles. 
Fig.~\ref{fig:dataDecon}(b) shows the joint distribution after the
application of the deconvolution algorithm. After this procedure, the negative
$\alpha$ artefacts disappear from the distribution. The method shows that there
is a cluster of semi-trapped particles ($\alpha\approx0.05$) which appears to
be distinctly separated from the rest of the sample. A second, less pronounced
subpopulation with $\alpha\approx0.35$ is also visible. Traces of these structures can be seen even without deconvolution, but with it they become much more pronounced, and we believe the reconstructed features are closer to the ground truth. \review{The limitation of this procedure is that it does not allow for multimodality tests as it returns a PDF rather than a sample. However, formal tests can be proposed given sufficiently simple hypotheses about the distribution of the studied sample. Alas, for this particular data set this distribution is likely complex.}

\section*{Discussion}

We presented a method to maximize the information gain from the TA-MSD of
individual diffusive trajectories. The proposed procedure enables a more
reliable estimation of the anomalous diffusion exponent $\alpha$ and the
diffusivity $D$ for anomalous diffusion processes \review{with stationary, Gaussian increments}. Since such data do not
fulfill the conditions necessary for a good performance of the classical
regression (the OLS method), the obtained estimates from this method are
both biased and have a large variance. The strength of both effects depends
on the window used for the fitting, the trajectory length, and the value of
the anomalous diffusion exponent. This makes finding and using an optimized
variant of the OLS method quite cumbersome and prone to substantial error.

The method we developed, i.e., the GLS method adapted to the TA-MSD analysis,
significantly improves the classical fitting procedure by
increasing accuracy, while maintaining an easy implementation and it can be automated. It works for data with experimental
noise present and even for ultra-short ($\approx10$ data points) trajectories,
for which extracting as much information as possible is even more important.
Furthermore, GLS is more efficient than OLS because it uses the entire TA-MSD data
and therefore does not ignore any part of the available information. The
GLS approach outlined here requires some preliminary estimate of $\alpha$. For
this purpose one can use prior experimental knowledge or the OLS estimate. In
the latter case our new approach serves as a correction and thus can be viewed
as a statistical enhancement of the classical method. 
Crucially, we provide formulas for the PDF of the errors of measured parameters,
so we can determine how reliable the estimation is. We consider these error estimates as
a major part of this work that will allow clear judgment on the reliability of
the fitted parameters for experimental measurements.

Furthermore, we can use this knowledge to reconstruct the original population structure of the data by using
the presented deconvolution procedure. By using the information contained in the entire set of
measured estimates this method can drastically reduce the blur introduced by
the estimation errors. The reconstructed density is sharper and much better reveals the differences between
different subpopulations of particles. This methodology can help to clearly distinguish what part of the parameter variability is an intrinsic property of the studied sample and what part is only a statistical artefact.

As next steps it is desirable to combine the GLS-deconvolution method with
complementary statistical observables such as the displacement correlation
function, the amplitude scatter PDF of the TA-MSD of a collection of SPT
trajectories, the mean-maximal excursion method or higher-order moments
\cite{tejedor,pccp,barkai2012strange,He08,kowalek2019classification}. In
particular, it could be interesting to develop the present approach further
for the mean-squared increment measure (or its higher order variants
\cite{kolmogorov1}) proposed recently \cite{wei,wei1}.
Another possible route to explore is the use of the GLS method in the
determination of change-points in the observed motion, as an application
of its ability to identify subpopulations. It will be interesting to see
how this approach competes with existing change-point analysis methods \cite{andi2,henrik2, kucharczyk18, malinowski25}. \review{ Knowledge about the statistical properties of the TA-MSD can also help in quantifying the effects of discretization: the differences between using discrete data and estimators, and using continuous models and results based on them.}

The main limitation of our approach is that the method requires the knowledge
on how TA-MSD errors behave, which is model-dependent. Hence, one must assume
a diffusion model describing the data behavior. In the idealized situation
when the entire distribution of the TA-MSD errors is known, the GLS is the
best possible estimate. Our tests prove that for systems such as mixtures of
FBMs the loss of power due to the lack of knowledge is surprisingly small
and the method is still very efficient. The knowledge of the stochastic
model underlying SPT data can be obtained from machine-learning assisted
model classification methods \cite{henrik,henrik1,andi2,munoz2020single,
muozGil2021,kowalek2019classification}.
Of course, in a realistic setting the anomalous diffusion modeling can have
a vast range of possible complexities which need to be considered case by
case. However, the GLS approach makes such considerations explicit, whereas
the classical OLS approach essentially uses a model which never really fits
the diffusion data. As such, it does not guarantee any controllable efficiency.
To conclude we point out that the proposed approach is quite flexible and has
good potential to be extended for other types of diffusion processes. 
Taking into account what statistical errors are expected in the physical model
under consideration and how will it affect the data analysis is still a too rarely discussed aspect of experimental science.

\section*{Author Contributions}

JŚ and JJ created the statistical procedures. JŚ performed the simulations and made the GitHub package. DK provided the experimental data. DK and RM supervised the biophysical analysis. All authors wrote the manuscript.

\section*{Acknowledgements}

RM acknowledges the German Science Foundation (DFG, grant nos. ME 1535/22-1
and INST 336/234-1 within CRC 1294). DK acknowledges the support of the National Science Foundation (NSF) Grant 2102832.

\section*{Published version}

The published version of this article is available at \href{https://doi.org/10.1016/j.bpj.2026.07.008}{https://doi.org/10.1016/j.bpj.2026.07.008}.






\bibliography{bibliography_cleared}



\appendix
\renewcommand{\thefigure}{S\arabic{figure}}
\renewcommand{\thesection}{S\arabic{section}}
\renewcommand{\thetable}{S\arabic{table}}
\renewcommand{\theequation}{S\arabic{equation}}
\setcounter{equation}{0}

\section{Regression for TA-MSD}

We use the following basic notation:
\begin{itemize}
	\item $n$, $m$, $d$, $b$ denote the trajectory length, number of TA-MSD points
	used, trajectory dimension, logarithm base
	\item $\bd x$, $\bd y$, $\bd\epsilon$ or $\bd e$, $\bd\beta$: vector of log
	time labels, vector of log TA-MSD values, vectors of errors (estimate minus
	true value), vector of estimates of $\log(2dD)$ and $\alpha$;
	\item $\widehat{x}$: estimate of the quantity $x$;
	\item $\langle\bd\cdot\rangle_{(D,\alpha)}$: averaging with $(D,\alpha)$ fixed,
	i.e., the conditional expectation $\E[\bd\cdot|D,\alpha]$, analogically for $\langle\bd\cdot\rangle_{\widehat\alpha}$.
\end{itemize}

\subsection{MSD in linear scale}
\label{s:err}

Given a measured trajectory, the expected errors of the TA-MSD
Eq.~2 are $\epsilon_k\coloneqq\overline{\delta^2_k}-\langle\delta^2
_k\rangle\da$, where the subpopulation MSD $\langle\delta^2_k\rangle\da$ is
essentially determined by the local parameters of the traced particle. The
TA-MSD estimator is unbiased, $\langle\epsilon_k\rangle\da=0$. We are interested
in the error distribution primarily characterized by the covariance matrix
$\langle\epsilon_k\epsilon_l\rangle_{(D,\alpha)}$, which is a function of 4th
order moments of the values $X_i$. The exact and general formula can be given
in the case of the trajectory-wise Gaussian system, i.e., when the trajectories
are Gaussian with possibly varying parameters. Then, the error covariance depends
on the covariance of the increments $X_{i+k}-X_i$, which we denote by
\begin{equation}
	c_{i,j,k,l}\coloneqq\langle(X_{i+k}-X_i)(X_{j+l}-X_j)\rangle\da=r_{i,j}+r_{i+k,j+l}
	-r_{i,j+l}-r_{i+k,j},\quad c_k\coloneqq c_{0,0,k,k}=\langle\delta^2_k\rangle\da,
\end{equation}
where $r$ is the trajectory-wise covariance function of $X$, $r_{i,j}=\langle X_i
X_j\rangle\da$. For the one-dimensional FBM model used later this reads $r_{i,j}
=D(\Delta t)^\alpha(i^\alpha+j^\alpha-|i-j|^\alpha)$; depending on the concrete
choice an additional factor $1/2$ might be present and/or the so-called Hurst
index $H=\alpha/2$ is used.

The necessary 4th-order moments can then be calculated using the Wick-Isserlis
theorem, yielding
\begin{equation}
	\label{eq:msdCovSum}
	\langle\epsilon_k\epsilon_l\rangle\da=\frac{2}{(n-k)(n-l)}\sum_{i=1}^{n-k}\sum_{
		j=1}^{n-l}(c_{i,j,k,l})^2.
\end{equation}
For a $d$-dimensional motion with independent coordinates the total errors add
up, so this formula should be multiplied by the number of dimensions $d$.

For complicated covariance structures, such as FBM with $\alpha\neq1$, there is
no simple analytical expression for this sum, and it requires $4(n-k)(n-l)$
covariance computations for each $(k,l)$. Its numerical calculation can be
vastly fastened if we notice that this matrix is symmetric (so only half of
the terms needs to be calculated) and $c_{i,j,k,l}$ is a function of $j-i$
only. The number of repeated terms is one plus the lengths of the $45^\circ$
lines crossing the rectangle $(n-k)\times(n-l)$. The reduced formula can be
written as
\begin{equation}
	\label{eq:reduced}
	\sum_{i=1}^{n-k}\sum_{j=1}^{n-l}(c_{i,j,k,l})^2=\sum_{h=2}^{n-l}(c_{1,h,k,l})^2
	(n-l-h+1)+\sum_{h=1}^{n-k}(c_{h,1,k,l})^2\begin{cases}n-l,& h\le l-k+1,\\
		n-k-h+1,& h>l-k+1,\end{cases}
\end{equation}
where $k\le l$. The matrix is symmetric so that values for $k>l$ are obtained by
switching the order of $k,l$.

\subsection{MSD in log-log scale} 

Establishing the general behavior of the TA-MSD logarithm is a quite
complicated problem, because the logarithm is a highly non-linear function.
It can be studied numerically using Monte Carlo methods or expressed as a
Gaussian integral. However, for longer trajectories and small errors
analytical results are also available.

\begin{figure}
	\centering
	\includegraphics[width=0.5\linewidth]{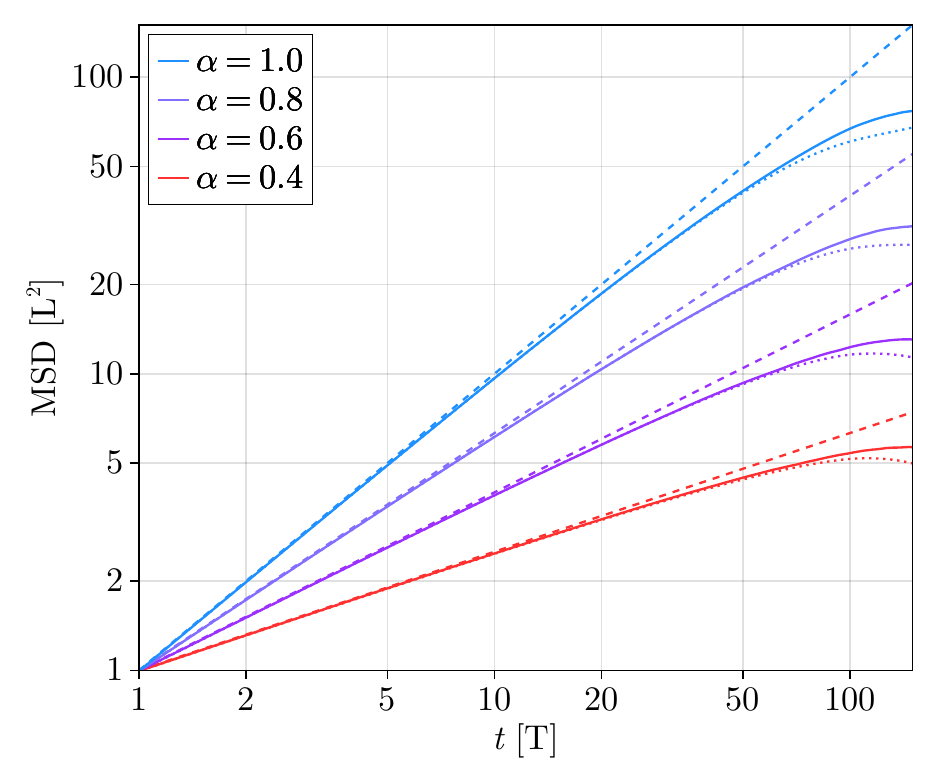}
	\caption{Visualization of log bias for the TA-MSD calculated from FBM
		trajectories with 200 points, simulated in numerical units with $D=1/2$ and $\Delta t=1$. What is shown
		are the averages calculated in the log space, as those affect the fitting. Solid line is average of the simulated log TA-MSD. Dashed line is the exact  log MSD, i.e. $\alpha \log_{10}t +\log_{10}(2D) $. Dotted line is the average of log TA-MSD  corrected by the factor $-\ln b\Sigma_{k,k}/2$. Sample size was $10^4$ trajectories.}
	\label{fig:bias}
\end{figure}

To obtain these analytical expressions, we expand the logarithm into a Taylor
series around the true value of the MSD, $\log_b\overline{\delta^2_k}=\log_b
\langle\delta^2_k\rangle\da+\epsilon_k/\big(\ln b\langle\delta^2_k \rangle\da
\big)-\epsilon_k^2/\big(2\ln b\langle\delta_k^2\rangle\da^2\big)+\mathcal{O}(
\epsilon_k^3)$, where $\mathcal{O}$ is the Landau symbol. The estimator $\log_b
\overline{\delta^2_k}$ is therefore biased, $\langle\log_b\overline{\delta^2_k}
-\log_b\delta^2_k\rangle\da\approx-\langle\epsilon_k^2\rangle\da/\big(2\ln b
\langle\delta_k^2\rangle\da^2\big)$. Its variance is dominated by the term
$\epsilon_k/\big(\ln b\langle\delta^2_k\rangle\da\big)$, so that the deviations
from the mean $e_k\coloneqq\log_b\overline{\delta^2_k}-\log_b\langle\delta^2_k
\rangle\da$ have approximately the TA-MSD error covariance matrix
\ref{eq:msdCovSum}, but divided by a prefactor $\ln b$ and MSD values. We
will denote the log MSD error covariance by $\Sigma$. Explicitly it is given
in the form
\begin{equation}
	\label{eq:sigma}
	\Sigma_{k,l}\coloneqq\langle e_ke_l\rangle\da\approx\frac{1}{(\ln b)^2c_kc_l}
	\frac{2}{(n-k)(n-l)}\sum_{i=1}^{n-k}\sum_{j=1}^{n-l}(c_{i,j,k,l})^2.
\end{equation}
For the TA-MSD of a $d$-dimensional motion with independent components the
numerator is multiplied by $d$ and the numerator by $d^2$, so overall it is
divided by $d$. This matrix does not depend on $D$, which greatly simplifies
the analysis of the log TA-MSD. Conveniently, the approximate bias can be
expressed as $-\ln b\Sigma_{k,k}/2$. In a typical setting this approximation
works quite well, see Fig.~\ref{fig:bias}. This plot also shows that for FBM
the bias is a negative concave function whose amplitude increases with the
errors. It means that without correction the estimator of $\alpha$ (slope) has
a negative bias and that of $D$ (intercept) has a positive bias. For each
trajectory length those increase with the size of the TA-MSD window used for
the fit.

How well this covariance and bias approximations work depends on the amplitude
of the normalized noise $\epsilon_k/\langle\delta_k^2\rangle\da$. Numerical
tests show that it is sufficient even for reasonably short trajectories of
around 50 points or more.

As a last note, one should not use more terms in the Taylor expansion of the
logarithm in the hope of a better approximation. Such additional terms diverge
from the original function more outside of the closest neighbourhood of $\log_b
\langle\delta_k^2\rangle\da$, and then the fraction of medium to large errors
$\epsilon_k/\langle\delta_k^2\rangle\da\gtrsim1$ distorts the result more,
not less.

\subsection{Classical regression}
\label{s:OLS}

Fitting a straight line to a set of points is perhaps one of the most prominent
problems in the whole field of statistics. Here we start with the standard
approach used for the TA-MSD. To simplify the formulas we denote the $m$ plotted
points ($m<n$) as $x_k=\log(k\Delta t),y_k=\log\overline{\delta^2_k}$, their
intercept as $\beta_1=\log2dD$, the slope as $\beta_2=\alpha$, and write the
regression problem in both expanded and vector forms\footnote{The problem as
	given is two-dimensional ($\bd\beta$ has 2 entries and $\bd x$ has two columns),
	although the methods we use work for any dimension.}
\begin{equation}
	y_k=\beta_2x_k+\beta_1+e_k\quad\mathrm{or}\quad\bd y=\bd x\bd\beta+\bd e,
\end{equation}
with
\begin{equation}
	\bd\beta=\begin{bmatrix}\beta_1\\\beta_2 \end{bmatrix},\quad\bd y=\begin{bmatrix}
		y_1\\y_2\\\vdots\\y_m\end{bmatrix},\quad\bd x=\begin{bmatrix}1& x_1\\1& x_2\\
		\vdots&\vdots\\1& x_m\end{bmatrix},\quad\bd e=\begin{bmatrix}e_1\\e_2\\\vdots\\
		e_m\end{bmatrix}.
\end{equation}
Importantly, the so-called design matrix $\bd x$ is fixed and we assume that
its measurement uncertainties are negligible. This is valid in typical SPT
experiments, where possible errors in measuring time are not relevant when
compared to the measured displacements.

Now, the simplest and most common approach is to look for a line which is
overall the closest to the observed points, minimizing the sum $\sum_{i=1}^m
(y_i-\beta_2x_i-\beta_1)^2=\Vert\bd y-\bd x\bd\beta\Vert^2$. This is the
ordinary least squares (OLS) method yielding the estimator
\begin{equation}
	\label{eq:OLS}
	\widehat{\bd\beta}_\mathrm{OLS}=(\bd x^\T\bd x)^{-1}\bd x^\T\bd y.
\end{equation}
Here $(\bd x^\T\bd x)^{-1}\bd x^\T$ is the $2\times m$ response matrix which
maps the observed $\bd y$ to the desired parameters. Substituting $\bd y=\bd
x\bd\beta+\bd e$ shows that the expected OLS errors are $\bd e_\mathrm{OLS}=
\widehat{\bd\beta}_\mathrm{OLS}-\bd\beta=(\bd x^\T\bd x)^{-1}\bd x^\T\bd e$.
Consequently, they are unbiased, $\langle\bd e_\mathrm{OLS}\rangle_{(D,\alpha)}
=0$, if $\langle\bd e\rangle=0$, and their expected covariance is
\begin{equation}
	\label{eq:olscov}
	\langle\bd e_\mathrm{OLS}\bd e_\mathrm{OLS}^\T\rangle\da=(\bd x^\T\bd x)^{-1}
	\bd x^\T\langle\bd e \bd e^\T\rangle_{(D,\alpha)}\bd x(\bd x^\T\bd x)^{-1}=(
	\bd x^\T\bd x)^{-1}\bd x^\T\Sigma\bd x(\bd x^\T \bd x)^{-1}.
\end{equation}
The errors are also expected to have an approximate normal distribution for sufficient trajectory length \cite{linMod}.
As shown in Sec.~\ref{s:err}, after fitting the data, the trajectory-wise
log TA-MSD error matrix $\Sigma$ can also be estimated. The obtained error
matrix of the estimator is a function of the design matrix and the true
values of $D$ and $\alpha$.

It is known that for iid errors with $\Sigma=\sigma_\mathrm{err}I$, where $I$
is the identity matrix, the OLS method provided the best possible estimate. We
can treat this case as a rough approximation when nothing about the errors is
known. In such a case the error 
global variance $\sigma^2_\mathrm{err}$ can be estimated from the residuals
and used to approximate the parameter covariance,
\begin{equation}
	\label{eq:OLSrough}
	\langle\bd e_\mathrm{OLS}\bd e_\mathrm{OLS}^\T\rangle\da\approx\sigma_\mathrm{
		err}(\bd x^\T\bd x)^{-1}=\frac{\sigma_\mathrm{err}}{m\sum_{i=1}^mx_i^2-\left(
		\sum_{i=1}^mx_i\right)^2}\begin{bmatrix}\sum_{i=1}^mx_i^2&-\sum_{i=1}^m x_i\\
		-\sum_{i=1}^mx_i&m
	\end{bmatrix}.
\end{equation}
In particular, the sign of the covariance $\langle e_1 e_2\rangle\da$ depends
on $-\sum_{i=1}^m x_i$. For large $\Delta t$ and $x_k=\log(k\Delta t)$ this
covariance is expected to be negative, for small $\Delta t$ to be positive.
This can serve as a rough expectation of how the plot of $(\log \widehat{D},
\widehat{\alpha})$ is stretched due to the estimation errors.

While the OLS estimator has a nice geometrical interpretation, for the TA-MSD
the iid assumption is severely broken and the OLS estimation becomes rather
inefficient. One essentially important consequence of this problem is that
the OLS becomes worse when applied to too many TA-MSD values, and for short
trajectories the loss quickly becomes drastic. The intuitive explanation is
that the TA-MSD values calculated at larger times are calculated from a smaller
amount of data (as the number of longer available increments decreases) and
are less statistically reliable, but the OLS cannot account for that. Choosing
the optimal TA-MSD window to use in OLS requires complicated calibration
curves \cite{kepten2015guide}. However, this deficiency can be corrected.

\subsection{Generalized regression}
\label{s:GLS}

One way to obtain an efficient estimation of the TA-MSD with realistic
assumptions is to use the maximum likelihood method. We intend to maximize
the probability of parameters given observations, $p(\bd\beta|\bd y)$. From
Bayes' rule
\begin{equation}
	p(\bd\beta|\bd y)=\frac{p(\bd y|\bd\beta)p(\bd\beta)}{p(\bd y)}.
\end{equation}
The factor $p(\bd y)$ does not matter, as it does not depend on $\bd\beta$.
Without prior knowledge, one can take $p(\bd\beta)=1$. The last term $p(\bd
y|\bd\beta)$ can be approximated as Gaussian with mean $\log\delta^2_k=\bd x
\bd\beta$ and the covariance being the log MSD error covariance $\Sigma$. The
problem becomes equivalent to finding the maximum of the Gaussian PDF $p(\bd
\beta|\bd y)\propto\exp(-\frac{1}{2}(\bd y-\bd x\bd\beta)^\T\Sigma^{-1}(\bd y
-\bd x\bd \beta))$, yielding the estimate
\begin{equation}
	\widehat{\bd\beta}=\mathop{\mathrm{argmin}}_{\bd\beta}\,(\bd y-\bd x\bd\beta)
	^\T\Sigma^{-1}(\bd y-\bd x\bd\beta).
\end{equation}
As a quadratic form, this problem can be solved using standard methods, with
the optimum being achieved for
\begin{equation}
	\label{eq:gls}
	\widehat{\bd\beta}_\mathrm{GLS}=(\bd x^\T\Sigma^{-1}\bd x)^{-1}\bd x^\T\Sigma^{
		-1}\bd y.
\end{equation}
Note that the form of the generalized least squares (GLS) estimator is similar
to the OLS. In fact it can be interpreted as the OLS in the new coordinates
$\widetilde{\bd y}\coloneqq\sqrt{\Sigma^{-1}}\bd y,\widetilde{\bd x}\coloneqq\sqrt{
	\Sigma^{-1}}\bd x$. These new linearly transformed variables also fulfill the
relation $\widetilde{\bd y}=\widetilde{\bd x}\bd\beta+\widetilde{\bd e}$, but
the transformed errors $\widetilde{\bd e}\coloneqq\sqrt{\Sigma^{-1}}\bd e$ are
iid. Hence, the OLS is optimal for the transformed coordinates, and consequently
GLS is optimal for the original coordinates. In a (typical) situation when we
do not know $\Sigma$ exactly, we want to approximate it as well as possible.
Nevertheless, if $\widetilde{\bd e}$ are closer to being iid than ${\bd e}$,
the estimation should be enhanced.

Averaging over the equation $\bd x^\T\Sigma^{-1}\bd x\widehat{\bd\beta}
_\mathrm{GLS}=\bd x^\T\Sigma^{-1}\bd y$ shows that, similarly to OLS, the
obtained estimator is unbiased, $\langle\widehat{\bd\beta}_\mathrm{GLS}\rangle
\da=\bd\beta$ as long as $\bd y$ is unbiased. Importantly, being unbiased
does not depend on the used matrix $\Sigma$. For an incorrect matrix, the
estimation errors will not be optimal, but it will still be unbiased. As
expected, assuming iid errors $\Sigma\propto I$, GLS simplifies to OLS.

Like for OLS, the GLS errors are also expected to be approximately normally
distributed \cite{linMod}. Their covariance can be calculated in the same way
as for OLS, yielding the surprisingly simple formula
\begin{equation}
	\label{eq:glscov}
	\langle\bd e_\mathrm{GLS}\bd e_\mathrm{GLS}^\T\rangle\da=(\bd x^\T\Sigma^{-1}
	\bd x)^{-1}.
\end{equation}
This result holds if $\Sigma$ is the exact error covariance matrix and, in
this case, GLS is guaranteed to be optimal (i.e., the variances in relation
\ref{eq:glscov} are a lower bound for all possible fits) and better than
OLS. However, $\Sigma$ is a function of the estimated parameters $\log D,
\alpha$. We can use a four-step procedure: 
\begin{itemize}
	\item[(i)]Use a potentially inefficient method (e.g., OLS) to get a
	preliminary estimate $\widehat{\bd\beta}{}^\mathrm{init}$.
	\item[(ii)] Use $\widehat{\alpha}$ to approximate the TA-MSD error covariance
	$\widehat{\Sigma}^\mathrm{init}$.
	\item[(iii)] Subtract the log TA-MSD bias $\approx-\ln b\widehat{\Sigma}
	^\mathrm{init}_{k,k}/2$ from $\bd y$.
	\item[(iv)] Substitute $\widehat{\Sigma}^\mathrm{init}$ into Eq.~\ref{eq:gls}
	to obtain the final enhanced GLS estimate.
\end{itemize}
Such an approach is an example of \emph{feasible\/} GLS \cite{econometrics}. Obviously,
we lose some power as compared to the situation when $\Sigma$ is known perfectly.
However, because the variability of $\widehat{\Sigma}^\mathrm{init}$ is limited
due to our strong assumptions about the model (for FBM it is a function of
$\alpha$ only), the loss is small {as shown in 
Fig.~2. Given the assumption about $\widehat{\Sigma}_\mathrm{
	init}$ we can assess theoretically how much loss can be expected using
\begin{equation}
	\label{eq:glscovInit}
	\langle\bd e_\mathrm{GLS}\bd e_\mathrm{GLS}^\T\rangle\da=\big(\bd x^\T(
	\widehat{\Sigma}^\mathrm{init})^{-1}\bd x\big)^{-1}\bd x^\T(\widehat{
		\Sigma}^\mathrm{init})^{-1}\Sigma(\widehat\Sigma^\mathrm{init})^{-1}\bd x\big(
	\bd x^\T(\widehat{\Sigma}^\mathrm{init})^{-1}\bd x\big)^{-1}.
\end{equation}
{In Fig.~3}} we show an example of the improvement in the estimation results if the GLS method is used instead of OLS. The main disadvantage of OLS is that because of the broken assumptions the estimation gets derailed for too many TA-MSD values used. 

\subsection{Asymptotic Gaussianity of the estimation errors}
\label{s:gauss}

For Gaussian dynamics like FBM or fractional Langevin equation the TA-MSD has a generalized $\chi^2$ distribution, being a sum of squares of Gaussian variables \cite{sikoraTest}. Obtaining the features of this distribution requires diagonalising the highly-dimensional covariance matrix, a complicated procedure \cite{grebenkov2011,sikoraTest}. After calculating logarithms and linearly transforming them to obtain anomalous diffusion estimates the analytical form of their distribution becomes extremely complex.

Fortunately, even for relatively short trajectories using the asymptotic distribution is viable. The setting is as follows: given fixed estimation window we are interested in the distribution of vector $\overline{\delta^2_1}, \overline{\delta^2_2}, \ldots, \overline{\delta^2_w} $ as the trajectory length $n$ increases. This question was thoroughly studied by Didier and Zhang \cite{didier2017}. The outline of the argument is that the squared increments inherit the dependence from the underlying motion. If it is not too strong, central limit theorem predicts that their averaged sums (that is TA-MSD) become asymptotically Gaussian. For the FBM and similar motions like fractional Langevin dynamics the critical $\alpha$ is 3/2. Below that the limit is Gaussian and above that it is much more complicated multivariate Rosenblatt-type distribution \cite{didier2017}.

As for large $n$ this distribution becomes concentrated around its mean $\langle\delta^2_1\rangle,\langle\delta^2_2\rangle,\ldots,\langle\delta^2_w\rangle$ the transformed log TA-MSD is dominated by the first order expansion around it,  $\log_b\overline{\delta^2_k}\approx\log_b
\langle\delta^2_k\rangle\da+\epsilon_k/\big(\ln b\langle\delta^2_k \rangle\da
\big)$ and is also asymptotically Gaussian (for $\alpha < 3/2$); this argument is called the delta method \cite{doob1935}. Notice that it is precisely the expansion that we used to calculate the covariance and bias of the log TA-MSD. The OLS and GLS are linear transformations of the log TA-MSD, so they are Gaussian as well with covariances \ref{eq:olscov} and \ref{eq:glscov}. Small point to note is that the feasible GLS might not be Gaussian if it uses a random preliminary estimate, but if it is a consistent estimate, the Gaussianity will be preserved, though the convergence rate will be slower.

\begin{figure*}[]
	\centering
	\begin{tikzpicture}
		\draw(0,0) node[inner sep=0]{\includegraphics[width=0.9\textwidth]{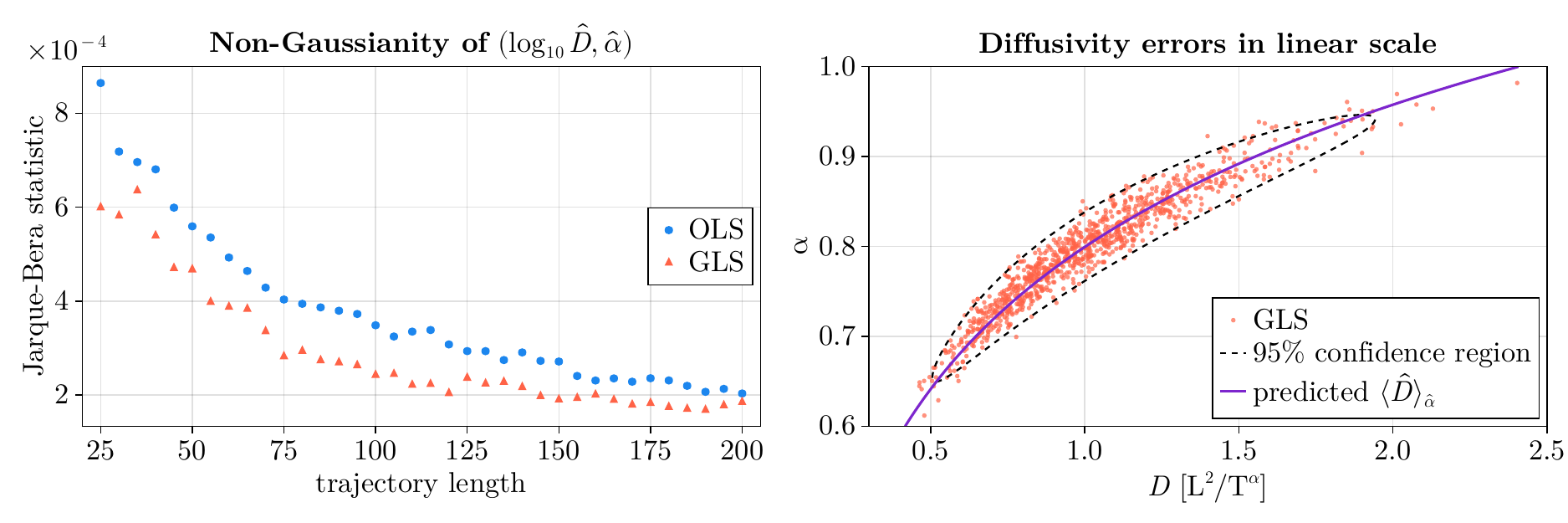}};
		\draw(-8.0,2.3) node {(a)};
		\draw(0.0,2.3) node {(b)};
	\end{tikzpicture}
	\caption{(a) Jarque-Bera non-Gaussianity measure for OLS and GLS estimates estimated from averaging 10 times samples of $10^5$ trajectories. Both estimates exhibit high-level of Gaussianity even calculated using short trajectories. (b) The distribution of $(\widehat{D},\widehat{\alpha})$ is approximately Gaussian for $\alpha$ and log-Gaussian for $D$. As a consequence exponential formula for the conditional expectancy Eq. \ref{eq:condD} (violet line) holds. The model was 2D FBM with true $D=1, \alpha = 0.8, \Delta t =  0.01$. In (b) the trajectory length was 100; $10^4$ points was shown. }\label{fig:gaussianity}
\end{figure*}

In Fig. \ref{fig:gaussianity} (a) we check if sufficient convergence to a Gaussian is reached for reasonably sized trajectories. We quantify it using 2-dimensional Jarque-Bera statistic divided by the sample size which measures how much 3rd and 4th moments are far from those of a Gaussian \cite{doornik}. For a comparison 2D Laplace distribution would have this statistic equal to $3/4$; this proves that both OLS and GLS estimates are strongly Gaussian, even for short trajectories.

Having established the Gaussianity of $(\log_b \widehat D,\widehat \alpha)$, the pair $(\widehat D,\widehat\alpha)$ can be treated in two ways. By again using the delta method we know it is also asymptotically Gaussian. However, the function $b^{x}$ significantly thickens the right tail of $\widehat D$ compared to $\log_b \widehat D$ so that approximating $\widehat D$ as log Gaussian (log-normal) is more effective, at least for trajectories which are not very long. As a consequence, a particular feature can be observed: if we put $(\widehat D,\widehat \alpha)$ corresponding to a single true $(D,\alpha)$ on the scatterplot, the dependence visible on is on average exponential,
\begin{equation}\label{eq:condD}
	\langle \widehat D\rangle_{\widehat \alpha} = \frac{1}{2d}\e^{A(\widehat\alpha-\langle \widehat \alpha\rangle) + B}, \quad A = \ln b\frac{\cov(\widehat \beta_1,\widehat \alpha)}{\var\widehat \alpha}, B = \ln b\langle \widehat \beta_1\rangle + \frac{(\ln b)^2}{2}\left(\var \widehat \beta_1 - \frac{(\cov(\widehat\beta_1,\widehat\alpha))^2}{\var\widehat\alpha}\right),
\end{equation}
the required variances and covariances can be read from \ref{eq:olscov} and \ref{eq:glscov}.
This formula follows from the fact that $\widehat\beta_1 = \log_b(2d\widehat D)$ conditioned on $\widehat \alpha$ is also Gaussian and $\langle \widehat D\rangle_{\widehat \alpha}$ is just the moment generating function of this new variable. Such exponential dependence is observed in the single particle tracking experiments \cite{cherstvy2019non,benelli2021, korabel2021}. By the argument above this is a purely statistical effect caused by the regression applied to a population of single $(D,\alpha)$ trajectories which is unrelated to the underlying physical system. We illustrate this phenomenon in Fig. \ref{fig:gaussianity} (b) which shows that Eq. \ref{eq:condD} predicts the line on which the estimates are concentrated. More comprehensive prediction can be made by pushing 2D Gaussian confidence ellipsis from $\log D$ space to the linear space.

\subsection{Illustration: Ultra-short trajectories}
\label{s:short}
{As shown in Fig.~5, the GLS method improves the estimation for ultra-short trajectories. 
Here, we provide additional details. The error‑covariance approximation in Eq.~\ref{eq:sigma} loses accuracy for small number of data points. Therefore, instead of relying on the approximation, we compute the error covariance numerically. In this case, either Monte Carlo sampling or evaluation of the corresponding Gaussian integral can be used, and both approaches are computationally efficient for ultra‑short trajectories.}

\begin{table}[]\centering\small

\begin{tabular}{c|cccccc|cccccc|}
\cline{2-13}
                          & \multicolumn{6}{c|}{bias}                                                                                                                                                                                                           & \multicolumn{6}{c|}{variance}                                                                                                                                                                                                      \\ \cline{2-13} 
                          & \multicolumn{2}{c|}{OLS2}                                                          & \multicolumn{2}{c|}{GLS}                                                         & \multicolumn{2}{c|}{gain}                                    & \multicolumn{2}{c|}{OLS2}                                                      & \multicolumn{2}{c|}{GLS}                                                       & \multicolumn{2}{c|}{gain}                                    \\ \hline
\multicolumn{1}{|c|}{$\alpha$}  & \multicolumn{1}{c|}{\cellcolor{LightGrey}\scriptsize $\log_{10}\widehat D$}     & \multicolumn{1}{c|}{\footnotesize $\widehat{\alpha}$}      & \multicolumn{1}{c|}{\cellcolor{LightGrey}\scriptsize $\log_{10}\widehat D$}    & \multicolumn{1}{c|}{\footnotesize $\widehat{\alpha}$}      & \multicolumn{1}{c|}{\cellcolor{LightGrey}\scriptsize $\log_{10}\widehat D$}     & \footnotesize $\widehat \alpha$      & \multicolumn{1}{c|}{\cellcolor{LightGrey}\scriptsize $\log_{10}\widehat D$}   & \multicolumn{1}{c|}{\footnotesize $\widehat{\alpha}$}    & \multicolumn{1}{c|}{\cellcolor{LightGrey}\scriptsize $\log_{10}\widehat D$}   & \multicolumn{1}{c|}{\footnotesize$\widehat{\alpha}$}     & \multicolumn{1}{c|}{\cellcolor{LightGrey}\scriptsize $\log_{10}\widehat D$}     & \footnotesize $\widehat \alpha$      \\ \hline
\multicolumn{1}{|c|}{0.3} & \multicolumn{1}{c|}{\cellcolor{LightGrey}-0.048} & \multicolumn{1}{c|}{-0.014} & \multicolumn{1}{c|}{\cellcolor{LightGrey}0.057} & \multicolumn{1}{c|}{0.053}  & \multicolumn{1}{c|}{\cellcolor{LightGrey}-21\%}  & -290\% & \multicolumn{1}{c|}{\cellcolor{LightGrey}0.30}  & \multicolumn{1}{c|}{0.22} & \multicolumn{1}{c|}{\cellcolor{LightGrey}0.12} & \multicolumn{1}{c|}{0.09} & \multicolumn{1}{c|}{\cellcolor{LightGrey}61\%}   & 57\%   \\ \hline
\multicolumn{1}{|c|}{0.4} & \multicolumn{1}{c|}{\cellcolor{LightGrey}-0.050}  & \multicolumn{1}{c|}{-0.017} & \multicolumn{1}{c|}{\cellcolor{LightGrey}0.056} & \multicolumn{1}{c|}{0.051}  & \multicolumn{1}{c|}{\cellcolor{LightGrey}-12\%}  & -200\% & \multicolumn{1}{c|}{\cellcolor{LightGrey}0.29} & \multicolumn{1}{c|}{0.21} & \multicolumn{1}{c|}{\cellcolor{LightGrey}0.14} & \multicolumn{1}{c|}{0.10}   & \multicolumn{1}{c|}{\cellcolor{LightGrey}52\%}   & 50\%   \\ \hline
\multicolumn{1}{|c|}{0.5} & \multicolumn{1}{c|}{\cellcolor{LightGrey}-0.054} & \multicolumn{1}{c|}{-0.021} & \multicolumn{1}{c|}{\cellcolor{LightGrey}0.053} & \multicolumn{1}{c|}{0.047}  & \multicolumn{1}{c|}{\cellcolor{LightGrey}1\%} & -130\% & \multicolumn{1}{c|}{\cellcolor{LightGrey}0.28} & \multicolumn{1}{c|}{0.20}  & \multicolumn{1}{c|}{\cellcolor{LightGrey}0.16} & \multicolumn{1}{c|}{0.11}  & \multicolumn{1}{c|}{\cellcolor{LightGrey}44\%}   & 43\%   \\ \hline
\multicolumn{1}{|c|}{0.6} & \multicolumn{1}{c|}{\cellcolor{LightGrey}-0.059} & \multicolumn{1}{c|}{-0.025} & \multicolumn{1}{c|}{\cellcolor{LightGrey}0.050}  & \multicolumn{1}{c|}{0.044}  & \multicolumn{1}{c|}{\cellcolor{LightGrey}14\%}   & -73\%  & \multicolumn{1}{c|}{\cellcolor{LightGrey}0.28} & \multicolumn{1}{c|}{0.19} & \multicolumn{1}{c|}{\cellcolor{LightGrey}0.17} & \multicolumn{1}{c|}{0.12}  & \multicolumn{1}{c|}{\cellcolor{LightGrey}37\%}   & 36\%   \\ \hline
\multicolumn{1}{|c|}{0.7} & \multicolumn{1}{c|}{\cellcolor{LightGrey}-0.064} & \multicolumn{1}{c|}{-0.030}  & \multicolumn{1}{c|}{\cellcolor{LightGrey}0.047} & \multicolumn{1}{c|}{0.040}   & \multicolumn{1}{c|}{\cellcolor{LightGrey}26\%}   & -31\%  & \multicolumn{1}{c|}{\cellcolor{LightGrey}0.27} & \multicolumn{1}{c|}{0.18} & \multicolumn{1}{c|}{\cellcolor{LightGrey}0.19} & \multicolumn{1}{c|}{0.13}  & \multicolumn{1}{c|}{\cellcolor{LightGrey}29\%}   & 29\%   \\ \hline
\multicolumn{1}{|c|}{0.8} & \multicolumn{1}{c|}{\cellcolor{LightGrey}-0.071} & \multicolumn{1}{c|}{-0.036} & \multicolumn{1}{c|}{\cellcolor{LightGrey}0.043} & \multicolumn{1}{c|}{0.035}  & \multicolumn{1}{c|}{\cellcolor{LightGrey}39\%}   & 4.8\%  & \multicolumn{1}{c|}{\cellcolor{LightGrey}0.26} & \multicolumn{1}{c|}{0.17} & \multicolumn{1}{c|}{\cellcolor{LightGrey}0.21} & \multicolumn{1}{c|}{0.13}  & \multicolumn{1}{c|}{\cellcolor{LightGrey}22\%}   & 23\%   \\ \hline
\multicolumn{1}{|c|}{0.9} & \multicolumn{1}{c|}{\cellcolor{LightGrey}-0.076} & \multicolumn{1}{c|}{-0.041} & \multicolumn{1}{c|}{\cellcolor{LightGrey}0.040}  & \multicolumn{1}{c|}{0.031}  & \multicolumn{1}{c|}{\cellcolor{LightGrey}48\%}   & 26\%   & \multicolumn{1}{c|}{\cellcolor{LightGrey}0.25} & \multicolumn{1}{c|}{0.16} & \multicolumn{1}{c|}{\cellcolor{LightGrey}0.22} & \multicolumn{1}{c|}{0.13}  & \multicolumn{1}{c|}{\cellcolor{LightGrey}14\%}   & 17\%   \\ \hline
\multicolumn{1}{|c|}{1.0} & \multicolumn{1}{c|}{\cellcolor{LightGrey}-0.085} & \multicolumn{1}{c|}{-0.048} & \multicolumn{1}{c|}{\cellcolor{LightGrey}0.035} & \multicolumn{1}{c|}{0.025}  & \multicolumn{1}{c|}{\cellcolor{LightGrey}59\%}   & 49\%   & \multicolumn{1}{c|}{\cellcolor{LightGrey}0.25} & \multicolumn{1}{c|}{0.15} & \multicolumn{1}{c|}{\cellcolor{LightGrey}0.23} & \multicolumn{1}{c|}{0.13}  & \multicolumn{1}{c|}{\cellcolor{LightGrey}8\%}    & 11\%   \\ \hline
\multicolumn{1}{|c|}{1.1} & \multicolumn{1}{c|}{\cellcolor{LightGrey}-0.094} & \multicolumn{1}{c|}{-0.055} & \multicolumn{1}{c|}{\cellcolor{LightGrey}0.029} & \multicolumn{1}{c|}{0.019}  & \multicolumn{1}{c|}{\cellcolor{LightGrey}70\%}   & 66\%   & \multicolumn{1}{c|}{\cellcolor{LightGrey}0.25} & \multicolumn{1}{c|}{0.14} & \multicolumn{1}{c|}{\cellcolor{LightGrey}0.24} & \multicolumn{1}{c|}{0.13}  & \multicolumn{1}{c|}{\cellcolor{LightGrey}2.1\%}  & 6.3\%  \\ \hline
\multicolumn{1}{|c|}{1.2} & \multicolumn{1}{c|}{\cellcolor{LightGrey}-0.10}   & \multicolumn{1}{c|}{-0.063} & \multicolumn{1}{c|}{\cellcolor{LightGrey}0.024} & \multicolumn{1}{c|}{0.013}  & \multicolumn{1}{c|}{\cellcolor{LightGrey}77\%}   & 79\%   & \multicolumn{1}{c|}{\cellcolor{LightGrey}0.25} & \multicolumn{1}{c|}{0.13} & \multicolumn{1}{c|}{\cellcolor{LightGrey}0.25} & \multicolumn{1}{c|}{0.13}  & \multicolumn{1}{c|}{\cellcolor{LightGrey}-3.1\%} & 1.7\%  \\ \hline
\multicolumn{1}{|c|}{1.3} & \multicolumn{1}{c|}{\cellcolor{LightGrey}-0.12}  & \multicolumn{1}{c|}{-0.071} & \multicolumn{1}{c|}{\cellcolor{LightGrey}0.018} & \multicolumn{1}{c|}{0.0075} & \multicolumn{1}{c|}{\cellcolor{LightGrey}85\%}   & 89\%   & \multicolumn{1}{c|}{\cellcolor{LightGrey}0.25} & \multicolumn{1}{c|}{0.12} & \multicolumn{1}{c|}{\cellcolor{LightGrey}0.26} & \multicolumn{1}{c|}{0.12}  & \multicolumn{1}{c|}{\cellcolor{LightGrey}-7.5\%} & -1.9\% \\ \hline
\multicolumn{1}{|c|}{1.4} & \multicolumn{1}{c|}{\cellcolor{LightGrey}-0.13}  & \multicolumn{1}{c|}{-0.080}  & \multicolumn{1}{c|}{\cellcolor{LightGrey}0.011} & \multicolumn{1}{c|}{0.0010}  & \multicolumn{1}{c|}{\cellcolor{LightGrey}92\%}   & 99\%   & \multicolumn{1}{c|}{\cellcolor{LightGrey}0.25} & \multicolumn{1}{c|}{0.11} & \multicolumn{1}{c|}{\cellcolor{LightGrey}0.27} & \multicolumn{1}{c|}{0.11}  & \multicolumn{1}{c|}{\cellcolor{LightGrey}-11\%}  & -5\%   \\ \hline
\end{tabular}

	\vspace{\baselineskip}

\begin{tabular}{c|cccccc|cccccc|}
\cline{2-13}
                               & \multicolumn{6}{c|}{bias}                                                                                                                                                                                                                                                                                                                                                                           & \multicolumn{6}{c|}{variance}                                                                                                                                                                                                                                                                                                                                                                       \\ \cline{2-13} 
                               & \multicolumn{2}{c|}{OLS5}                                                                                                              & \multicolumn{2}{c|}{GLS}                                                                                                               & \multicolumn{2}{c|}{gain}                                                                                         & \multicolumn{2}{c|}{OLS5}                                                                                                              & \multicolumn{2}{c|}{GLS}                                                                                                               & \multicolumn{2}{c|}{gain}                                                                                         \\ \hline
\multicolumn{1}{|c|}{$\alpha$} & \multicolumn{1}{c|}{\cellcolor{LightGrey}\scriptsize $\log_{10}\widehat D$} & \multicolumn{1}{c|}{\footnotesize$\widehat \alpha$} & \multicolumn{1}{c|}{\cellcolor{LightGrey}\scriptsize $\log_{10}\widehat D$} & \multicolumn{1}{c|}{\footnotesize$\widehat \alpha$} & \multicolumn{1}{c|}{\cellcolor{LightGrey}\scriptsize $\log_{10}\widehat D$} & \footnotesize$\widehat \alpha$ & \multicolumn{1}{c|}{\cellcolor{LightGrey}\scriptsize $\log_{10}\widehat D$} & \multicolumn{1}{c|}{\footnotesize$\widehat \alpha$} & \multicolumn{1}{c|}{\cellcolor{LightGrey}\scriptsize $\log_{10}\widehat D$} & \multicolumn{1}{c|}{\footnotesize$\widehat \alpha$} & \multicolumn{1}{c|}{\cellcolor{LightGrey}\scriptsize $\log_{10}\widehat D$} &\footnotesize $\widehat \alpha$ \\ \hline
\multicolumn{1}{|c|}{0.3}      & \multicolumn{1}{c|}{\cellcolor{LightGrey}-0.069}                                           & \multicolumn{1}{c|}{-0.033}            & \multicolumn{1}{c|}{\cellcolor{LightGrey}-0.004}                                           & \multicolumn{1}{c|}{-0.005}           & \multicolumn{1}{c|}{\cellcolor{LightGrey}94\%}                                             & 86\%              & \multicolumn{1}{c|}{\cellcolor{LightGrey}0.11}                                             & \multicolumn{1}{c|}{0.09}              & \multicolumn{1}{c|}{\cellcolor{LightGrey}0.10}                                             & \multicolumn{1}{c|}{0.09}             & \multicolumn{1}{c|}{\cellcolor{LightGrey}8.7\%}                                            & 8.8\%             \\ \hline
\multicolumn{1}{|c|}{0.4}      & \multicolumn{1}{c|}{\cellcolor{LightGrey}-0.079}                                           & \multicolumn{1}{c|}{-0.043}            & \multicolumn{1}{c|}{\cellcolor{LightGrey}-0.011}                                           & \multicolumn{1}{c|}{-0.011}            & \multicolumn{1}{c|}{\cellcolor{LightGrey}87\%}                                             & 74\%              & \multicolumn{1}{c|}{\cellcolor{LightGrey}0.13}                                             & \multicolumn{1}{c|}{0.10}              & \multicolumn{1}{c|}{\cellcolor{LightGrey}0.13}                                             & \multicolumn{1}{c|}{0.10}             & \multicolumn{1}{c|}{\cellcolor{LightGrey}6.1\%}                                            & 6.9\%             \\ \hline
\multicolumn{1}{|c|}{0.5}      & \multicolumn{1}{c|}{\cellcolor{LightGrey}-0.091}                                           & \multicolumn{1}{c|}{-0.055}            & \multicolumn{1}{c|}{\cellcolor{LightGrey}-0.018}                                           & \multicolumn{1}{c|}{-0.018}            & \multicolumn{1}{c|}{\cellcolor{LightGrey}80\%}                                             & 67\%              & \multicolumn{1}{c|}{\cellcolor{LightGrey}0.15}                                             & \multicolumn{1}{c|}{0.11}              & \multicolumn{1}{c|}{\cellcolor{LightGrey}0.15}                                             & \multicolumn{1}{c|}{0.11}              & \multicolumn{1}{c|}{\cellcolor{LightGrey}4.0\%}                                            & 5.5\%             \\ \hline
\multicolumn{1}{|c|}{0.6}      & \multicolumn{1}{c|}{\cellcolor{LightGrey}-0.10}                                            & \multicolumn{1}{c|}{-0.065}            & \multicolumn{1}{c|}{\cellcolor{LightGrey}-0.024}                                           & \multicolumn{1}{c|}{-0.024}            & \multicolumn{1}{c|}{\cellcolor{LightGrey}76\%}                                             & 64\%              & \multicolumn{1}{c|}{\cellcolor{LightGrey}0.17}                                             & \multicolumn{1}{c|}{0.12}              & \multicolumn{1}{c|}{\cellcolor{LightGrey}0.17}                                             & \multicolumn{1}{c|}{0.12}              & \multicolumn{1}{c|}{\cellcolor{LightGrey}2.7\%}                                            & 4.8\%             \\ \hline
\multicolumn{1}{|c|}{0.7}      & \multicolumn{1}{c|}{\cellcolor{LightGrey}-0.11}                                            & \multicolumn{1}{c|}{-0.075}            & \multicolumn{1}{c|}{\cellcolor{LightGrey}-0.030}                                            & \multicolumn{1}{c|}{-0.028}            & \multicolumn{1}{c|}{\cellcolor{LightGrey}74\%}                                             & 62\%              & \multicolumn{1}{c|}{\cellcolor{LightGrey}0.19}                                             & \multicolumn{1}{c|}{0.13}              & \multicolumn{1}{c|}{\cellcolor{LightGrey}0.19}                                             & \multicolumn{1}{c|}{0.13}              & \multicolumn{1}{c|}{\cellcolor{LightGrey}1.8\%}                                            & 4.6\%             \\ \hline
\multicolumn{1}{|c|}{0.8}      & \multicolumn{1}{c|}{\cellcolor{LightGrey}-0.13}                                            & \multicolumn{1}{c|}{-0.087}            & \multicolumn{1}{c|}{\cellcolor{LightGrey}-0.036}                                           & \multicolumn{1}{c|}{-0.034}            & \multicolumn{1}{c|}{\cellcolor{LightGrey}71\%}                                             & 61\%              & \multicolumn{1}{c|}{\cellcolor{LightGrey}0.21}                                             & \multicolumn{1}{c|}{0.14}              & \multicolumn{1}{c|}{\cellcolor{LightGrey}0.20}                                             & \multicolumn{1}{c|}{0.13}              & \multicolumn{1}{c|}{\cellcolor{LightGrey}1.1\%}                                            & 4.5\%             \\ \hline
\multicolumn{1}{|c|}{0.9}      & \multicolumn{1}{c|}{\cellcolor{LightGrey}-0.14}                                            & \multicolumn{1}{c|}{-0.098}            & \multicolumn{1}{c|}{\cellcolor{LightGrey}-0.042}                                           & \multicolumn{1}{c|}{-0.039}            & \multicolumn{1}{c|}{\cellcolor{LightGrey}70\%}                                             & 61\%              & \multicolumn{1}{c|}{\cellcolor{LightGrey}0.22}                                             & \multicolumn{1}{c|}{0.15}              & \multicolumn{1}{c|}{\cellcolor{LightGrey}0.22}                                             & \multicolumn{1}{c|}{0.14}              & \multicolumn{1}{c|}{\cellcolor{LightGrey}0.78\%}                                           & 4.9\%             \\ \hline
\multicolumn{1}{|c|}{1.0}      & \multicolumn{1}{c|}{\cellcolor{LightGrey}-0.15}                                            & \multicolumn{1}{c|}{-0.11}             & \multicolumn{1}{c|}{\cellcolor{LightGrey}-0.048}                                           & \multicolumn{1}{c|}{-0.043}            & \multicolumn{1}{c|}{\cellcolor{LightGrey}69\%}                                             & 61\%              & \multicolumn{1}{c|}{\cellcolor{LightGrey}0.24}                                             & \multicolumn{1}{c|}{0.15}              & \multicolumn{1}{c|}{\cellcolor{LightGrey}0.24}                                             & \multicolumn{1}{c|}{0.14}              & \multicolumn{1}{c|}{\cellcolor{LightGrey}0.65\%}                                           & 5.7\%             \\ \hline
\multicolumn{1}{|c|}{1.1}      & \multicolumn{1}{c|}{\cellcolor{LightGrey}-0.16}                                            & \multicolumn{1}{c|}{-0.12}             & \multicolumn{1}{c|}{\cellcolor{LightGrey}-0.051}                                           & \multicolumn{1}{c|}{-0.045}            & \multicolumn{1}{c|}{\cellcolor{LightGrey}69\%}                                             & 62\%              & \multicolumn{1}{c|}{\cellcolor{LightGrey}0.25}                                             & \multicolumn{1}{c|}{0.15}              & \multicolumn{1}{c|}{\cellcolor{LightGrey}0.25}                                             & \multicolumn{1}{c|}{0.14}              & \multicolumn{1}{c|}{\cellcolor{LightGrey}0.65\%}                                           & 6.7\%             \\ \hline
\multicolumn{1}{|c|}{1.2}      & \multicolumn{1}{c|}{\cellcolor{LightGrey}-0.18}                                            & \multicolumn{1}{c|}{-0.13}             & \multicolumn{1}{c|}{\cellcolor{LightGrey}-0.056}                                           & \multicolumn{1}{c|}{-0.048}            & \multicolumn{1}{c|}{\cellcolor{LightGrey}69\%}                                             & 63\%              & \multicolumn{1}{c|}{\cellcolor{LightGrey}0.27}                                             & \multicolumn{1}{c|}{0.15}              & \multicolumn{1}{c|}{\cellcolor{LightGrey}0.27}                                             & \multicolumn{1}{c|}{0.14}              & \multicolumn{1}{c|}{\cellcolor{LightGrey}0.79\%}                                           & 8.0\%             \\ \hline
\multicolumn{1}{|c|}{1.3}      & \multicolumn{1}{c|}{\cellcolor{LightGrey}-0.19}                                            & \multicolumn{1}{c|}{-0.14}             & \multicolumn{1}{c|}{\cellcolor{LightGrey}-0.059}                                           & \multicolumn{1}{c|}{-0.050}            & \multicolumn{1}{c|}{\cellcolor{LightGrey}69\%}                                             & 64\%              & \multicolumn{1}{c|}{\cellcolor{LightGrey}0.28}                                             & \multicolumn{1}{c|}{0.15}              & \multicolumn{1}{c|}{\cellcolor{LightGrey}0.28}                                             & \multicolumn{1}{c|}{0.14}              & \multicolumn{1}{c|}{\cellcolor{LightGrey}0.82\%}                                           & 9.5\%             \\ \hline
\multicolumn{1}{|c|}{1.4}      & \multicolumn{1}{c|}{\cellcolor{LightGrey}-0.21}                                            & \multicolumn{1}{c|}{-0.15}             & \multicolumn{1}{c|}{\cellcolor{LightGrey}-0.064}                                           & \multicolumn{1}{c|}{-0.052}            & \multicolumn{1}{c|}{\cellcolor{LightGrey}69\%}                                             & 65\%              & \multicolumn{1}{c|}{\cellcolor{LightGrey}0.30}                                             & \multicolumn{1}{c|}{0.15}              & \multicolumn{1}{c|}{\cellcolor{LightGrey}0.29}                                             & \multicolumn{1}{c|}{0.13}              & \multicolumn{1}{c|}{\cellcolor{LightGrey}0.96\%}                                           & 11\%            \\ \hline
\end{tabular}

	\caption{Comparison between OLS applied to two (upper table) or five values 		of the TA-MSD (bottom table) and its GLS corrections for ultra-short, 10-point FBM trajectories.  The values in the "gain" columns are calculated as $100\% 		\cdot(1-(\mbox{GLS result})/(\mbox{OLS result}))$; \review{its positive values indicate better performance of the GLS method}.  For two points of the
		TA-MSD the OLS method has a large variance, but a small bias for sub- and
		normal diffusion. For subdiffusion the GLS significantly reduces the variance;
		in some cases it has a worse bias, but looking at its values $\approx0.05$ it
		is still fairly low. For superdiffusion GLS yields a slightly higher variance,
		but greatly corrects the bias, up to even 80 times. OLS applied to five values
		of the TA-MSD has a much lower variance than in the case of 2-TA-MSD values,
		but at the cost of bias increase. GLS strongly corrects the bias and further
		decreases the variance, making the estimation better for all $\alpha$. Sample size was $10^5$ trajectories.}
	\label{tab:short}
\end{table}

In Table~\ref{tab:short} we compare entire trajectory GLS to 2-point OLS and 5-point OLS. The results show that OLS applied to 2-point TA-MSD values should not be used for subdiffusion due to the resulting large variance. However, in terms of variance using such window becomes the best choice for superdiffusion, likely because long trends specific for long memory skew the OLS estimation for larger windows. Using
OLS for five TA-MSD values results in a lower variance for subdiffusion, but
also in an unnecessary bias caused by the log skew of larger errors at the
end of the window. In each of these cases, the GLS method achieves further
improvement, drastically removing larger values of the bias and reducing
the variance. The optimal procedure becomes: for subdiffusion use a medium
number of TA-MSD values in the OLS and improve it with GLS; for
superdiffusion use a small number of TA-MSD values in the OLS and also
improve it with GLS.

In many realistic scenarios the system represents a mixture of trajectories
that are both sub- and superdiffusive. In such a situation there is no good
choice of the number of TA-MSD points in OLS. However, using 5-point OLS with GLS improvement results in a good estimation quality for all $\alpha$, with only fairly small loses for superdiffusion.

The results presented are "the worst case possible" in that we only assume
that $\alpha>0.1$ and blindly use a rough initial OLS estimate, which has a
low quality. If we expect that the system is, for example, subdiffusive or
normal-diffusive with $\alpha\le1$ and use this condition in the GLS, it
can get better from few to several percent in gain. The OLS approach cannot
include such additional information.

\review{
\subsection{Illustration: Generalized Langevin equation}
\label{s:GLE}
}
Majority of this work is concentrated  on the FBM as it exbibits pure (and therefore the simplest to analyse) power law MSD. Still, we want to demonstrate the GLS is also suitable for other Gaussian processes with stationary increments. Among non-FBM models the most essential for biophysical applications is the generalized Langevin equation with power law kernel.

We consider a free motion GLE which is an equation for the particle's velocity $V(t)$ \cite{kouPRL, kouAAS}
\begin{equation}
    m\frac{\mathrm d}{\mathrm d t} V(t) = -\zeta\int_{-\infty}^t\mathrm d s \ V(s)K_\gamma(t-s)+ \sqrt{2\zeta k_BT}\frac{\mathrm d}{\mathrm d t} B_\gamma(t),
\end{equation}
with the kernel $K_\gamma$ equal to the covariance fractional noise force $\mathrm dB_\gamma(t)/\mathrm d t$ due to fluctuation-dissipation theorem, $K_\gamma(t) = \gamma(\gamma-1)|t|^{\gamma-2}$. We are interested in stationary (that is: long time) solutions, which have a simple form in the Fourier space. In particular, the covariance of the velocity is
\begin{equation} \label{eq:GLEsol} 
    \langle V(u)V(u+t)\rangle = \frac{1}{2\pi}\int_{-\infty}^\infty \mathrm d\omega\ \e^{\mathrm i t\omega} f(\omega),\quad f(\omega)\coloneqq \frac{2k_BT\zeta \Gamma(\gamma+1)\sin(\gamma\pi/2)|\omega|^{1-\gamma}}{\left|\zeta\Gamma(\gamma+1)|\omega|^{1-\gamma}(\sin(\gamma\pi/2)-\mathrm i\cos(\gamma\pi/2)\mathrm{sgn}(\omega)) -\mathrm im\omega\right|^2}.
\end{equation}
Integrating it twice we get the integral formulation of the displacement covariance
\begin{equation}
\langle X(s)X(t)\rangle = \frac{1}{\pi}\int_{0}^\infty \mathrm d\omega \left(\cos((t-s)\omega)-\cos(t\omega)-\cos(s\omega)+1\right) \omega^{-2}f(\omega)
\end{equation}
which we calculate numerically using Gaussian quadrature method \cite{numRec}. Having the covariance matrix, we simulate trajectories using Cholesky decomposition, the same as for the FBM. This approach guarantees being able to quickly compute a large number of short trajectories with minimal numerical errors.

As visible in Eq.~\ref{eq:GLEsol} the shape of the MSD depends on 2 parameters. The scale does not affect estimation, so in the simulations we fix $k_BT=1, m = 1$ and the two remaining parameters are $\gamma,\zeta$. At short times the motion is ballistic $\langle X(t)^2\rangle \propto t^2,t\to0$. At long times it is subdiffusive,
\begin{equation}
    \langle X(t)^2\rangle \sim k_B T\zeta^{-1} \frac{2\sin(\gamma\pi)}{\pi\gamma(1-\gamma)(2-\gamma)} t^{2-\gamma},\quad t\to\infty
\end{equation}
so fitting TA-MSD sufficiently far away from $t=0$ from the anomalous exponent we get $\widehat\gamma$ and from the diffusivity we get $\widehat\zeta$. The GLS procedure looks the same as for the FBM, the only difference being what covariance matrix is used. Parameters given by the OLS are used as preliminary estimates for the GLS, as always decision must be made what to do in the cases where these are nonsensical (that is $\widehat \zeta_{\mathrm{OLS}}<0$, $\widehat\gamma_{\mathrm{OLS}} < 1/2$,  $\widehat\gamma_{\mathrm{OLS}} > 1$; we used the average OLS values.

Exemplary results are shown in Fig. 6 of the main text.\\

\section{Advanced methods}

\subsection{Regression in the presence of additive noise}
\label{s:noise}

In many realistic settings experimental errors in SPT are non-negligible
during statistical analysis. Here we explain what to do in the presence
of simple, yet common, additive noise. Such "static noise" is typically
associated with the localization uncertainty, e.g. due to finite photon
counts or limitations to the optical setup itself \cite{meyer,berglund,savin}.
To account for such noise the observed data is assumed to have the form
$X_k^\mathrm{obs}=X_k^\mathrm{real}+\sigma \xi_k$ for any coordinate. Usually
the variance $\sigma^2$ is assumed to be constant for all trajectories and can
be determined beforehand (see also the discussion in \cite{meyer,jae_noisy}).
In this case the TA-MSD is not a power law, but has the offset form $\langle
\delta^2_k\rangle_{(D,\alpha)}=2dDt^{\alpha}+2d\sigma^2$. Hence, the log
TA-MSD is not a straight line and linear regression cannot be directly used.
Instead, one needs to use the corrected MSD $\log(\overline{\delta^2_k}-2d
\sigma^2)$, where the prefactor $2$ in $2d\sigma^2$ comes from the fact that
increments of white noise have twice larger variance than the original series.
One should be careful here, as this quantity may cause statistical problems if
$\sigma^2$ is comparable to the TA-MSD errors. This may happen especially at
short times as then $\overline{\delta^2_k}-2d\sigma^2$ will have mass near 0.

As before we need to determine the log errors covariance matrix, which can be
approximated as
\begin{equation}
	\Sigma_{k,l}^\mathrm{obs}\approx\frac{1}{(\ln b)^2}\frac{\langle\bd\epsilon
		_\mathrm{obs}\bd\epsilon_\mathrm{obs}^\T\rangle}{4d^2D^2\Delta t^{2\alpha}(kl)
		^{\alpha}}.
\end{equation}
The observed errors $\bd\epsilon_\mathrm{obs}$ correspond to the fourth central
moments of the position time series with noise. This quantity depends on $(D,
\alpha)$, which varies among trajectories, and the global $\sigma$. In this case
the dependence on $D$ does not cancel in the numerator and the denominator, and
the matrix $\Sigma_{k,l}^\mathrm{obs}$ depends on $\alpha$ and the ratio $\sigma/
D$. For each trajectory these two parameters must be used as a preliminary fit.
Similarly to the case without noise, one can tabularize error matrices for
$\alpha$s in the experimental range and the noise covariance matrix for $\sigma
=1$ to make the estimation faster.

For an efficient computation we must therefore separate the anomalous and the
white noise part. To this end, we need the fourth central moment, which is
twice the variance squared. We can decompose the error covariance into
\begin{equation}
	\langle\bd\epsilon_\mathrm{obs}\bd\epsilon_\mathrm{obs}^\T\rangle=\langle\bd
	\epsilon_\mathrm{real}\bd\epsilon_\mathrm{real}^\T\rangle+\sigma^2\bd S+
	\sigma^4\langle\bd\epsilon_\xi\bd\epsilon_\xi^\T\rangle,
\end{equation}
where $\bd S$ is a cross term matrix, i.e., the sum of products of the model
and the white noise covariance
\begin{equation}
	S_{k,l}=\frac{4}{(n-k)(n-l)}\sum_{i=1}^{n-k}\sum_{j=1}^{n-l}c_{i,j,k,l}(I_{i,j}
	+I_{i+k,j+l}-I_{i,j+l}-I_{i+k,j}).
\end{equation}
Here $I$ denotes the identity matrix. This sum is fast to compute because most
of the terms are zero. For the white noise itself we can determine the exact
TA-MSD error covariance. It can be done by counting the repeating $\xi_k$ in
sums for given $\epsilon_{\xi,k}$. The resulting matrix behaves differently on
the diagonal,
\begin{equation}
	\langle\epsilon_{\xi,k}\epsilon_{\xi,k}\rangle=\frac{4}{(n-k)^2}\begin{cases}
		3n-4k,&n\ge2k,\\2n-2k,&n< 2k,\end{cases}    
\end{equation}
and off-diagonal, for $l>k$
\begin{equation}
	\langle\epsilon_{\xi,k}\epsilon_{\xi,l}\rangle=\frac{4}{(n-k)(n-l)}\begin{cases}
		2n-k-2l,&n\ge k+l,\\n-l,&n<k+l.\end{cases}    
\end{equation}
Summarizing, one needs to compute the white noise error matrix once for $\sigma
=1$, then compute the error matrix of $\bd\epsilon_\mathrm{real}$ with $D=1$ and
cross term $\bd S$, both depending on $\alpha$. $\Sigma^\mathrm{obs}$ can then
be calculated for any trajectory using only rescaling and addition.

\begin{figure}
	\centering
	\begin{tikzpicture}
		\draw(0,0) node[inner sep=0]{\includegraphics[width=0.9\textwidth]{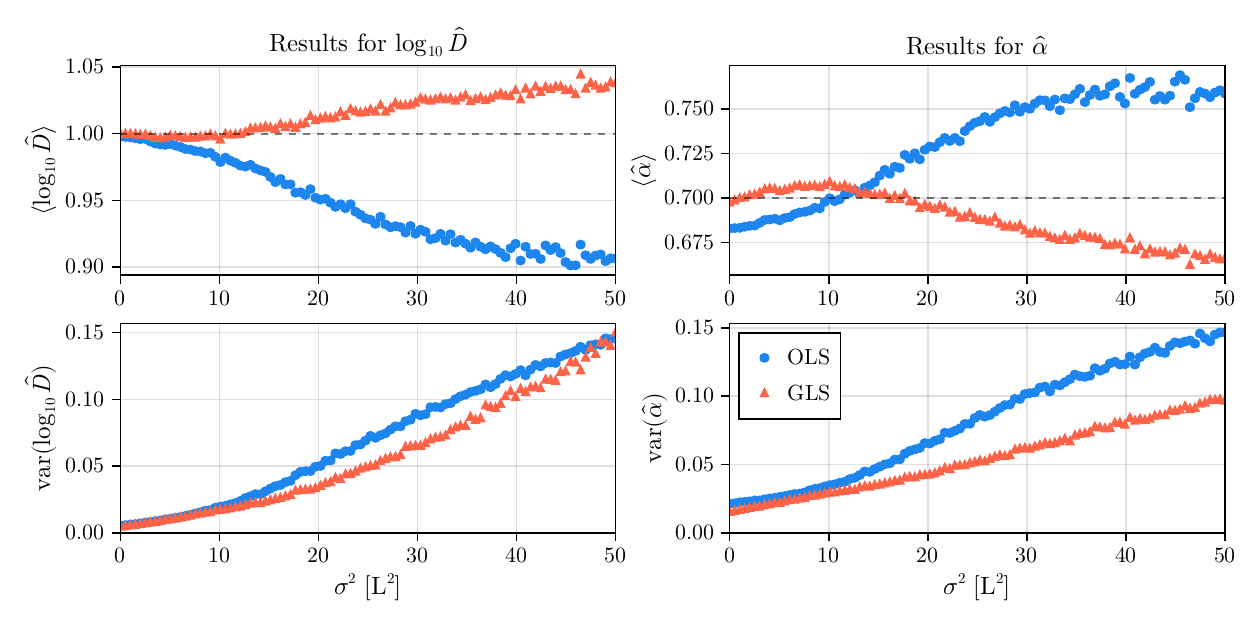}};
		\draw(-7.2,3.5) node {(a)};
		\draw(0.9,3.5) node {(b)};
		\draw(-7.2,0.2) node {(c)};
		\draw(0.9,0.2) node {(d)};
	\end{tikzpicture}
	
	\caption{Monte Carlo-calculated quality of the estimation (bias top, variance
		bottom) as a function of the experimental noise variance $\sigma^2$. The
		simulated trajectories were a one-dimensional FBM with $D=10$ and $\alpha=0.7$,
		with $100$ data points, in numerical units. OLS estimators were calculated
		using 10 points. (a,b) Biases of $\log_{10}\widehat D$ and $\widehat\alpha$. (c,d) Variances of $\log_{10}\widehat D$ and $\widehat\alpha$. Only trajectories with $0<\widehat{\alpha}<2$ were included
		due to some rare singular estimation values which would be removed in a
		practical setting. Sample size was $10^4$ trajectories.}
	\label{fig:noise}
\end{figure}
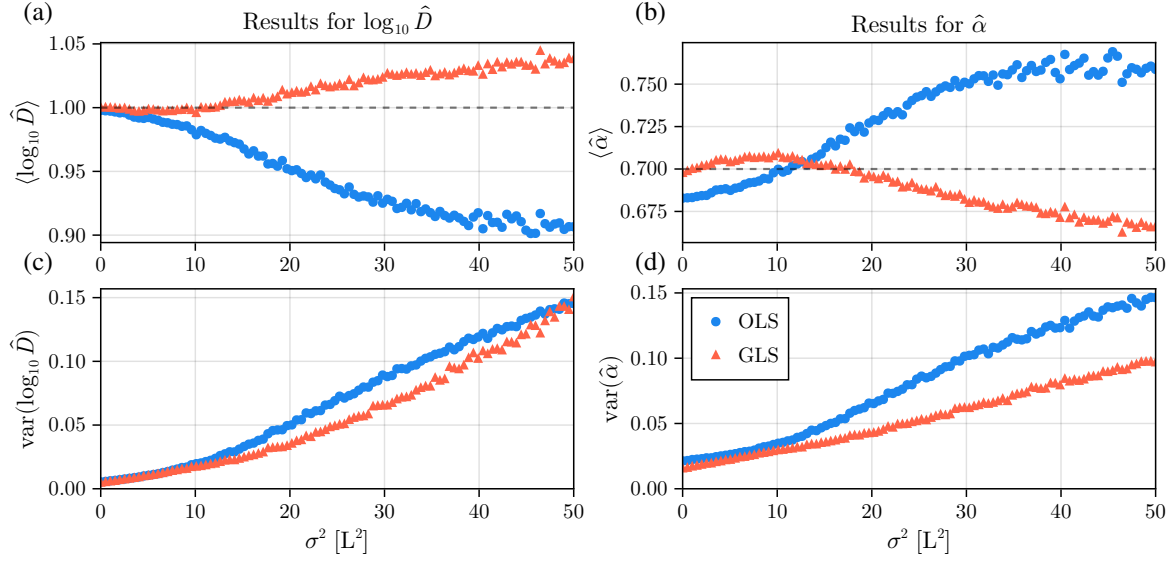

We check the efficiency of the method using numerical simulations of FBM. The
obtained results are plotted in Fig.~\ref{fig:noise}. We see that when the
variance of the noise reaches the value of the diffusivity (around $\sigma^2=
10$ on the plots), the OLS results start to deteriorate rapidly, both in bias
and variance. The GLS gains are very substantial, for $\alpha=0.7$ and high
$\sigma^2$ reducing the variance by a factor of two and the bias even by three
times. One flaw of the GLS method not visible on those plots is that for
$\sigma^2>25$ the number of trajectories with unphysical $\alpha<0$ or $\alpha
>2$ is slightly larger, around $8\%$ versus $5\%$ for the OLS method. Such
extreme values are also responsible for the quicker increase of the GLS
diffusivity variance. This is probably due to the fact that GLS uses more
data for which more problems with the logarithms of small values may appear;
these problems could be mitigated by a more careful TA-MSD post-processing.

\subsection{Split MSD estimators}
\label{sec:indep_est}

As can be seen from Eq.~\ref{eq:olscov}, there is an artificial dependence
between the estimators, stemming purely from the estimation method and not the
physical properties of the system. In order to avoid the occurrence of such
a dependence, one can estimate the parameters individually from disjoint
parts of a trajectory. Let $1<s<n$ be the split point, with the optimal value
being approximately in the middle, $s\approx(n-k)/2$. Then we define two
TA-MSDs calculated using only the first half (upper index $1/2$) and only the
second half (upper index $2/2$) of the trajectory,
\begin{equation}
	\label{eq:msdSplit}
	^{1/2}\overline{\delta^2_k}\coloneqq\frac{1}{s-k}\sum_{i=1}^{s-k}\big(X_{i+k}-
	X_i\big)^2,\quad ^{2/2}\overline{\delta^2_k}\coloneqq\frac{1}{n-s-k}\sum_{i=s+1}
	^{n-k}\big(X_{i+k}-X_i\big)^2.
\end{equation}
Based on the regression of those two parts, we obtain $^{1/2}\widehat{\bd
	\beta}$ and $^{2/2}\widehat{\bd\beta}$, respectively. As argued before, for the
best results the GLS method should be used. The mixed pairs $(\log^{1/2}\widehat{
	D},^{2/2}\widehat{\alpha})$ and $(\log^{2/2}\widehat{D},^{1/2}\widehat{\alpha})$
have significantly reduced correlation as compared to $(\log\widehat{D},\widehat{
	\alpha}$), as they depend on the noise from non-overlapping intervals, and only
the less significant dependence around the splitting point $s$ remains, i.e.,
\begin{equation}
	\label{eq:splitCov}
	\langle{}^{1/2}e_\mathrm{GLS}{}^{2/2}e_\mathrm{GLS}^\T\rangle\da=(\bd x^\T
	\Sigma^{-1}\bd x)^{-1}\bd x^\T\Sigma^{-1}\Sigma^\mathrm{split}\Sigma^{-1}\bd x
	(\bd x^\T\Sigma^{-1}\bd x)^{-1},
\end{equation}
with $\Sigma^\mathrm{split}$ being the cross-covariance of the corresponding
split log-errors,
\begin{equation}
	\label{sigsplit}
	\Sigma^\mathrm{split}_{k,l}=\frac{1}{(\ln b)^2 c_kc_l}\frac{2}{(n-s)(n-s-k)}
	\sum_{i=1}^{n-s}\sum_{j=s+1}^{n-k}(c_{i,j,k,l})^2
\end{equation}
for OLS, where as before we take $\Sigma\propto I$. Expression \ref{sigsplit}
for $\Sigma^\mathrm{split}$ is quite similar to Eq.~\ref{eq:sigma} for
$\Sigma$, the difference being that the $j$ indices are shifted by $+s$; the
data points are further away and their correlation is much smaller. The cost
of using this method is that by using twice shorter trajectories increases the
error amplitude \ref{eq:splitCov} as compared to the non-splitted estimator.

The significant advantage of this approach is that it is very straightforward:
it does not require any assumptions on the data nor multiple-step statistical
transformations.

\subsection{Illustration: estimating Brownian motion in different time windows}

As an illustration of slightly different setting let us imagine a situation in
which we want to fit the TA-MSD, but in an interval $[t_1,t_2]$ with $t_1>0$.
This may be the case, e.g., because the MSD values for short $t$ are more
affected by the noise, or because we want to separate distinct types of
dynamics at short and longer timescales.

We take the simplest example possible: Brownian motion in numerical units of time T and length L,
$t\in\{1,2,\ldots,200\}$. We attempt to fit the TA-MSD to the formula $2Dt+
\sigma$, where $\sigma$ may be due to static noise or some specific short-time
dynamics; the true values are $D=1/2$ and $\sigma=0$. Importantly, here in
contrast to Sec.~\ref{s:noise} we assume that we do not know $\sigma$, so both
parameters are to be estimated using TA-MSD regression.

\begin{figure}
	\centering
	\begin{tikzpicture}
		\draw(0,0) node[inner sep=0]{\includegraphics[width=0.777\textwidth]{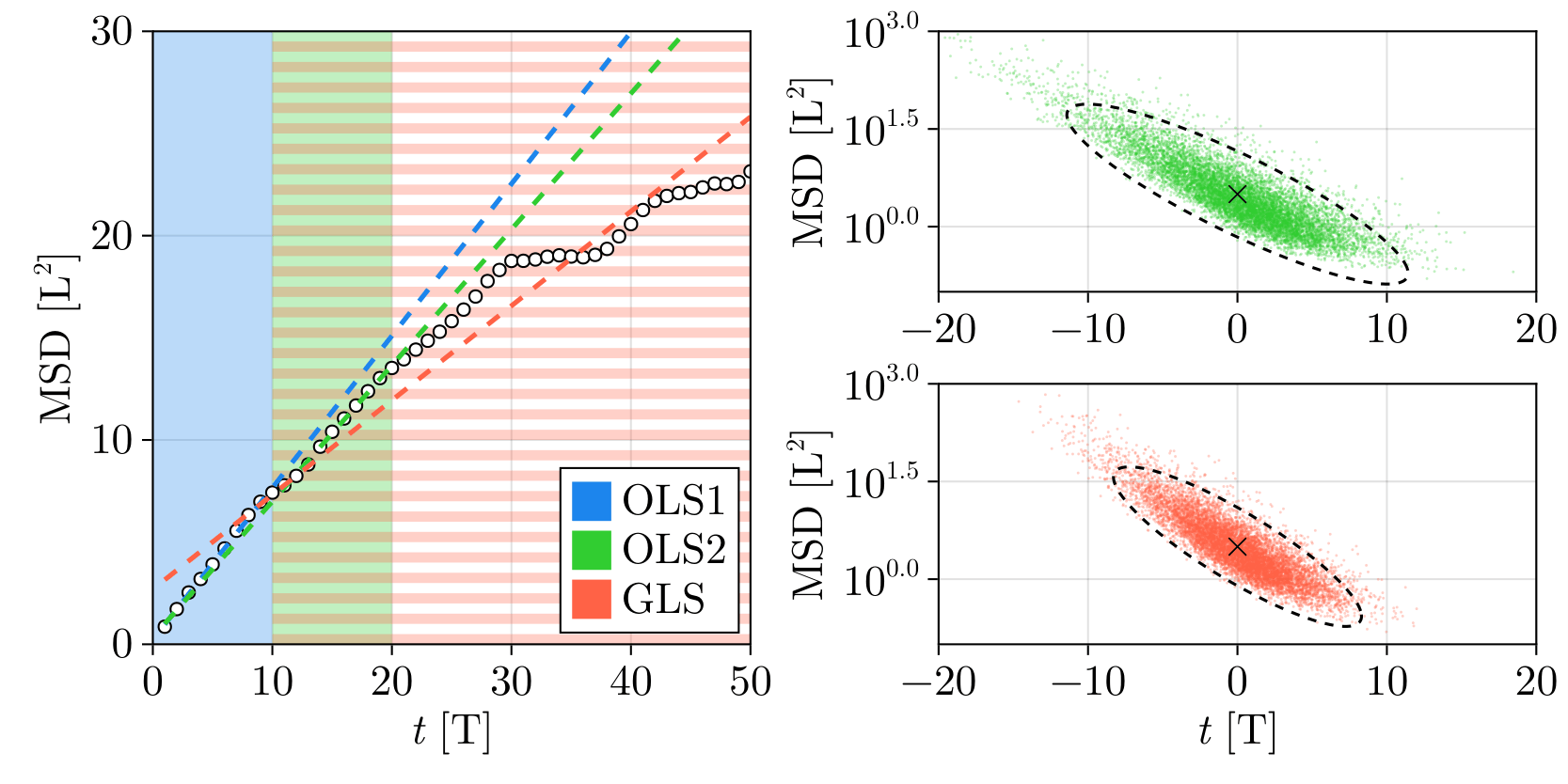}};
		\draw(-5.2,3.5) node {(a)};
		\draw(0.2,3.5) node {(b)};
		\draw(0.2,0.2) node {(c)};
	\end{tikzpicture}
	
	\caption{Fitting of the TA-MSD of a one-dimensional Brownian motion to the
		form $\delta^2(t)=2Dt+\sigma$ using different time intervals (in numerical
		units). (a) Exemplary trajectory with TA-MSD with fitting intervals (background) and fitted lines from the OLS1 (blue), OLS2 (green) and GLS (red strips). (b,c) Scatter plots of the estimated values together with the exact values (crosses) and theoretical error ellipses (dashed lines) for OLS2 and GLS. Sample size was $10^4$ trajectories.}
	\label{fig:interval}
\end{figure}
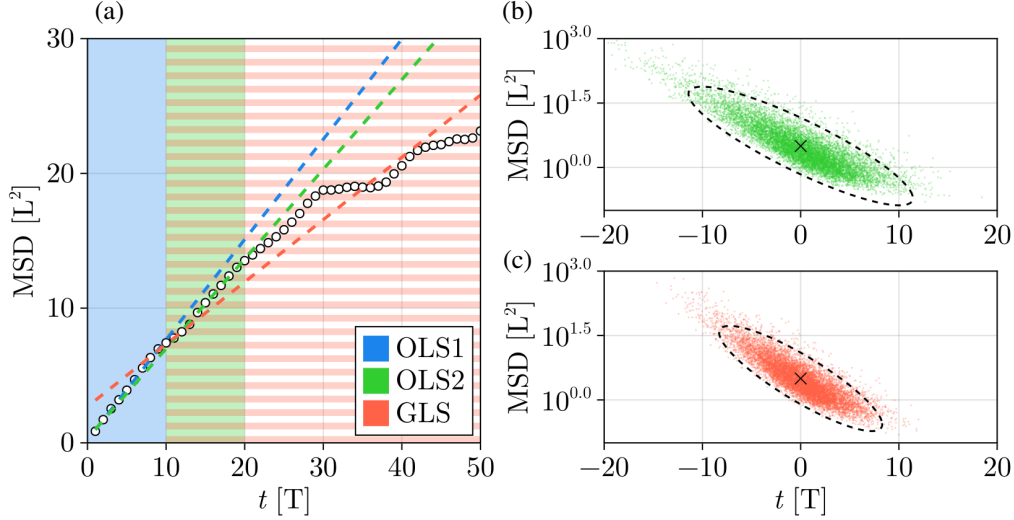

We use three estimators: OLS1 for the interval $\{1,12,\ldots,10\}$, OLS2 for
the interval $\{11,12,\ldots,20\}$, and GLS for the interval $\{11,12,\ldots,
200\}$. For GLS we assume the Brownian motion covariance, so the same error
matrix is used for all trajectories. The theoretical predictions for the
estimation covariances read
\begin{equation}
	\label{eq:interval}
	\langle\bd e_\mathrm{OLS1}\bd e_\mathrm{OLS1}^\T\rangle\approx\begin{bmatrix}
		0.35&-0.15\\-0.15&0.089\\\end{bmatrix},\quad
	\langle\bd e_\mathrm{OLS2}\bd e_\mathrm{OLS2}^\T\rangle\approx\begin{bmatrix}
		21.77&-2.31\\-2.31&0.32\\\end{bmatrix},\quad
	\langle\bd e_\mathrm{GLS}\bd e_\mathrm{GLS}^\T\rangle\approx\begin{bmatrix}
		11.50&-1.48\\-1.48&0.25\\\end{bmatrix},
\end{equation}
and we checked that they agree with the Monte Carlo -estimated covariances see Fig. \ref{fig:interval}.

The first point to draw our attention to is that it is natural to expect a
loss of efficiency in the intervals further away from $t=1$, but the 60-fold
increase of the variance $\widehat{\sigma}$, by moving only 10 points to the
right, is somewhat surprising. This observation proves that using our formulas
gives valuable insight, in this case showing the drastic unreliability of the
$\sigma$ estimation. One intuition behind this is that the intercept $\sigma$
is a crossing point between the lines $2Dt+\sigma$ and $t=1$, so errors in
determining the slope $2D$ add to it, multiplied by the factor $t_1=11$; this
is on top of having larger TA-MSD error variances. Using GLS decreases this
variance twice. All the estimators are also extremely correlated, the
covariances in Eq.~\ref{eq:interval} correspond to a correlation around $-88\%$.

The OLS2 error can be decreased by taking a smaller window---numerically calculating the parameter covariance matrices indicates that the optimum is only two points. In this case the response matrix is $(\bd x^\T\bd x)^{-1}\bd x^\T=[12\ -11; -1\ 1]$, which is nearly identical to the GLS response matrix, for which the rest of columns are close to zero. In this case one could guess the form of the estimator (though for general FBM it is not the case), but the fact that it is the optimal one is still non-trivial.

\section{Reconstruction of original parameters}

\subsection{Changing between log-log and normal scale}
\label{s:move}

In the previous Sections we left aside the fact that we estimate the regressors
$\bd\beta$, not $(D,\alpha)$ directly. The anomalous diffusion exponent is
estimated explicitly, but the relation between the diffusivity and the intercept
is non-linear, $\beta_1=\log_b(2dD)$, $D=b^{\beta_1}/2d$. This causes problems
in the sense that the fact that the estimator $\widehat{\beta}_1$ is unbiased
does not translate to an unbiased $\widehat{D}=b^{\widehat{\beta}_1}/2d$. In
Tab.~\ref{tab:DH} we show the trajectory-wise moments of this estimator, in the
first row for the expansion of small $\widehat{\bd\beta}-\bd\beta$, in the
second row in the limit of Gaussian $\widehat{\beta}$. The first order results
are obtained from the Taylor expansion of the function $b^x$. Results for
Gaussian variables come from the fact that for the Gaussian distribution all
the exponential averages of type $\langle b^X\rangle=\langle\e^{\ln b X}\rangle$
are elementary integrals which one can calculate exactly. Both agree up to the
first order in $\widehat{\bd\beta}-{\bd\beta}$, so the Gaussian approximation
is generally safe to use even if errors are not exactly Gaussian. The left
column of the table allows us to construct a corrected unbiased estimator,
which is $\widehat{D}_\mathrm{cor}\coloneqq\widehat{D}/\e^{w/2}$. For this new
estimator the variance and covariance (central and right columns) change,
they are simply divided by $\e^w$ or $\e^{w/2}$, respectively.

\begin{table}
	\centering
	\begin{tabular}{|l|c|c|c|}
		\hline
		& $\langle\widehat{D}\rangle\da$& $ \var(\widehat{D})\da$& $\cov(\widehat{D},
		\widehat{\alpha})\da$\\\hline
		first order & $D(1+w/2)$ & $D^2w$ & $cD$\\
		Gaussian & $D\e^{w/2}$ & $ D^2\e^w\left(\e^w-1\right) $ & $cD\e^w$\\\hline
	\end{tabular}
	\caption{Moments of $\widehat{D}=b^{\widehat{\beta}_1}/2$. Here $w\coloneqq
		(\ln b)^2\langle e_1^2\rangle\da$ and $c\coloneqq\frac{\ln b}{2}\langle e_1e_2
		\rangle\da$, these can be calculated from the error covariance formulas. The
		true diffusivity $D$ is unknown, but to estimate variance and covariance we
		can substitute $\widehat{D}$ or the corrected $\widehat{D}$.}
	\label{tab:DH}
\end{table}

\subsection{Joint density reconstruction}
\label{s:reconstruct}

Using Eqs.~\ref{eq:olscov} and \ref{eq:glscov} we can asses the regression
errors, so we know the approximate distribution of the $e$s for the observed
parameter values: $\alpha^\mathrm{obs}=\alpha^\mathrm{true}+e_\alpha$ and
$\log D^\mathrm{obs}=\log D^\mathrm{true}+e_{\log D}$. We aim at the
reconstruction of the properties of $(\log D^\mathrm{true},\alpha^\mathrm{true})$.
In this Section we work with $\log D$ instead of $D$. This is because $D$ is a scale
parameter, so in log scale it becomes an added constant. Consequently for the FBM model $e_{\log D}$ is completely independent of $D^\mathrm{true}$, being only a function
of $\alpha^\mathrm{true}$. Nonetheless, in a more complicated setting this may
no longer be true due to correlations of $D$ with other relevant parameters,
so the equation below is valid for the general case.

The joint error probability density is trajectory-wise dependent (in other
words: parameter dependent). Denote it as $p^\mathrm{err}_{e_{\log D},
	e_\alpha}(e_1,e_2|l_D,a)$, where $l_D$ and $a$ denote the values of the
parameters for a given trajectory. The observed distribution can be expressed
as the generalized convolution
\begin{equation}
	\label{eq:convolution}
	p^{\mathrm{obs}}_{\log D,\alpha}(l_D,a)=\int p^{\mathrm{true}}_{\log D,\alpha}
	(l_D',a')p^\mathrm{err}_{e_{\log D},e_\alpha}(l_D'-l_D,a'-a|l_D',a')\dd l_D'\dd
	a'.
\end{equation}
Crucially, in Sec. \ref{s:gauss} we have established that even for relatively short trajectories $p^\mathrm{err}_{e_{\log D},e_\alpha}(e_1,e_2|l_D',a')$ can be approximated by a Gaussian with known covariance. 

As one may guess, inverting the above formula in order to obtain $p^{\mathrm{
		true}}_{\log D,\alpha}$ from $p^{\mathrm{obs}}_{\log D, \alpha}$ may be
difficult. Such tasks are commonly performed during image analysis, where we
want to reconstruct a source image given a blurred picture. Here, the situation
is additionally complicated as the "blur" can and generally is position
dependent; moreover we do not know the exact $p^{\mathrm{obs}}$ but only the
estimate $\widehat p^{\;\mathrm{obs}}$. There are two ways to proceed: to make
strong assumptions about the errors or about the PDF. We will start with the
first approach.

\subsubsection{Non-parametric deconvolution}
\label{s:npDeconv}

As depicted in Fig.~2 the error covariance matrix does not
change rapidly in an experimentally realistic setting. At the cost of some
accuracy (which can be controlled) the problem can be simplified by assuming
the error distribution to be constant with respect to $(D,\alpha)$. A simple
and reasonable choice is to take the value of mean or median $\alpha$.
Alternatively, one can use the distribution averaged over the observed
$\alpha$. The second choice may catch more details of the sample, at the risk
of being more disturbed by the errors themselves. A constant error distribution
has the additional advantage that it is also statistically independent from the
parameters. Thus, the covariances add, and to reconstruct the true covariance
matrix of $(\log D,\alpha)$ we just subtract one of the errors.

For the PDF reconstruction the problem becomes a classical deconvolution, in
which we want to determine $p^\mathrm{true}$ given the convolution $p^\mathrm{
	obs}=p^\mathrm{true}*p^\mathrm{err}$. A simple division in Fourier space,
$\mathcal{F}[p^\mathrm{true}]=\mathcal{F}[p^\mathrm{obs}]/\mathcal{F}[p^\mathrm{
	err}]$ is not viable because it is numerically unstable. A better choice is the
Richardson--Lucy iterative algorithm, which is a general method commonly used in
image analysis. We construct a sequence $p^0=p^\mathrm{obs}$ (or some other
reasonable preliminary estimate) and then iterate on,
\begin{equation}
	p^{k+1}=p^k\cdot\left(\frac{p^\mathrm{obs}}{p^k*p^\mathrm{err}}*\widetilde{p}
	^{\,\mathrm{err}}\right)
\end{equation}
where $\widetilde{p}^{\,\mathrm{err}}$ is $p^\mathrm{err}$ with both arguments
reflected (Gaussian noise is symmetric, so in this case $\widetilde{p}^{\,
	\mathrm{err}}= p^\mathrm{err}$). This sequence converges to the desired
reconstruction, $p^k\xrightarrow{k\to\infty}p^\mathrm{true}$.

A more involved approach, which we call \textit{interpolated deconvolution},
uses the fact that we can repeat the above procedure using error estimates
for different $\alpha_i$. This sequence of PDFs can then be interpolated
creating an image for which the PDF values at each level $\alpha=\alpha_i$ are
calculated using $\alpha_i$ error estimate. This interpolation might not be
exactly normalized, so it should be suitably rescaled. The result is still
not the exact inverse of Eq.~\ref{eq:convolution}, but is a useful heuristic.

\begin{figure}[ht]
\centering
\begin{tikzpicture}
\draw(0,0) node[inner sep=0]{\includegraphics[width=1\textwidth]{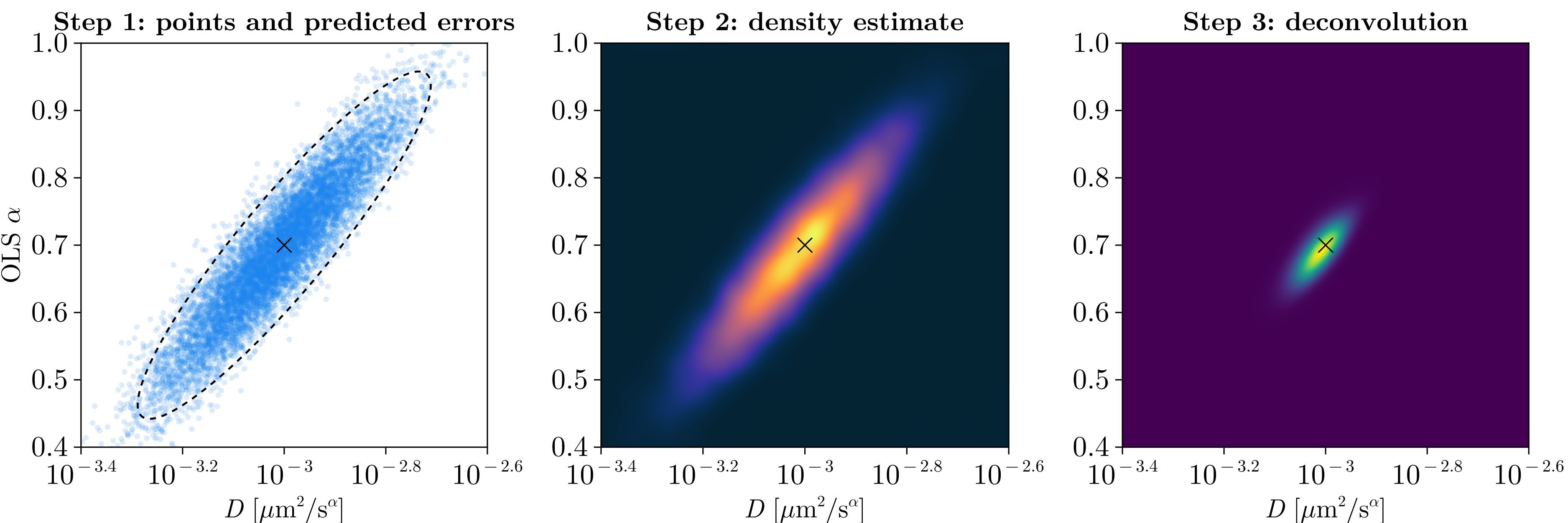}};
\draw(-8.6,2.65) node {(a)};
\draw(-2.0,2.65) node {(b)};
\draw(3.8,2.65) node {(c)};
\end{tikzpicture}
\caption{\review{Deconvolution applied to simulated FBM trajectories of length 100
with $\Delta t=0.0567$, $D=10^{-3}$ and $\alpha=0.7$ (marked by cross). The PDF was estimated with the kernel density method. (a) Scatter plots of OLS estimated values of $\alpha$ and $D$. Dashed line are the predicted 95\% confidence regions. (b) Joint PDFs of the estimates of $\alpha$ and $D$. (c)  Joint PDFs after deconvolution with the Richardson-Lucy algorithm.} Sample size was $10^4$ trajectories.} 
\label{fig:decEx2}
\end{figure}
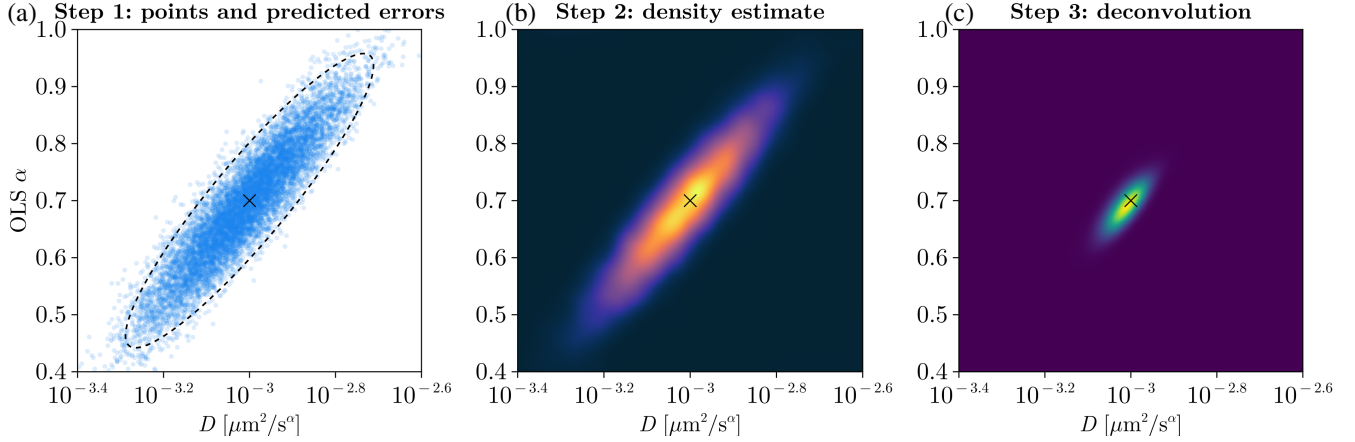

In the main text (Fig.~7) we show how a simple deconvolution
works for the single population of trajectories estimated with the GLS method, and the method proves to be very efficient. \review{Here in Fig. \ref{fig:decEx2} we complement it by also showing a deconvolution applied to the OLS estimates. The results are still good, but some accuracy was lost due to higher initial spread of the OLS.}

\begin{figure}
	\centering
	\begin{tikzpicture}
		\draw(0,0) node[inner sep=0]{\includegraphics[width=0.9\textwidth]{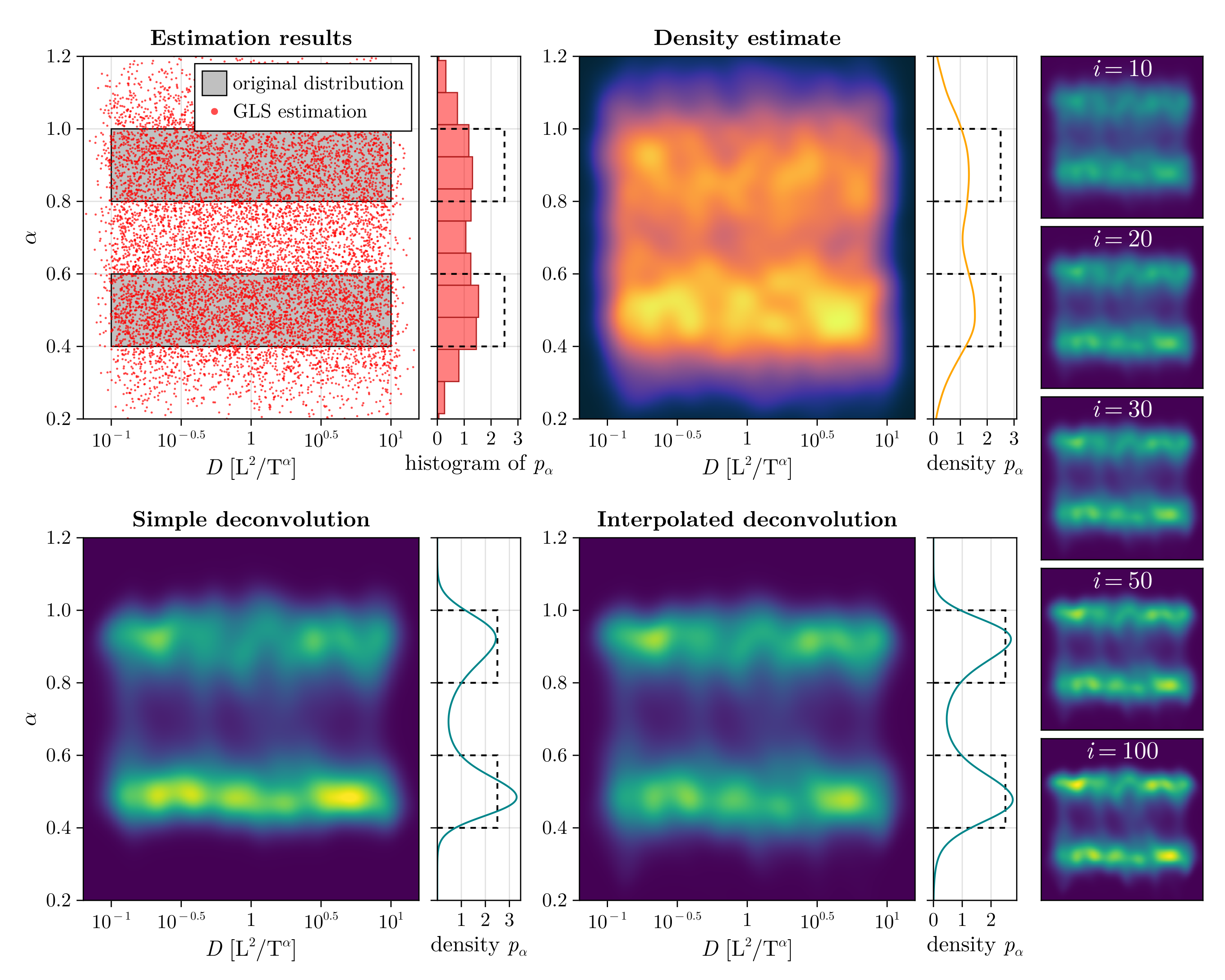}};
		\draw(-7.5,5.8) node {(a)};
		\draw(-1.1,5.8) node {(b)};
		\draw(-7.5,-0.4) node {(c)};
		\draw(-1.1,-0.4) node {(d)};
        \draw(6.5,5.8) node {(e)};
	\end{tikzpicture}
	
	\caption{Deconvolution procedure tested on a more complex original distribution:
		two rectangular populations. In numerical units. (a) Measured parameters and the original distribution of $(\log_{10} D,\alpha)$.  (b) The PDF was first approximated
		using a kernel density estimate, then, in (c) a simple deconvolution was performed
		assuming a constant error variance corresponding to $\alpha=0.7$. The
		interpolated deconvolution approach in (d) used values of $\alpha$ corresponding to
		each point of the PDF. \review{ (e) Dependence on the Richardson-Lucy algorithm iteration number $i$ for the interpolated deconvolution. (c) and (d) correspond to $i=50$.} Sample size was $10^4$ trajectories.}
	\label{fig:deconTest}
\end{figure}
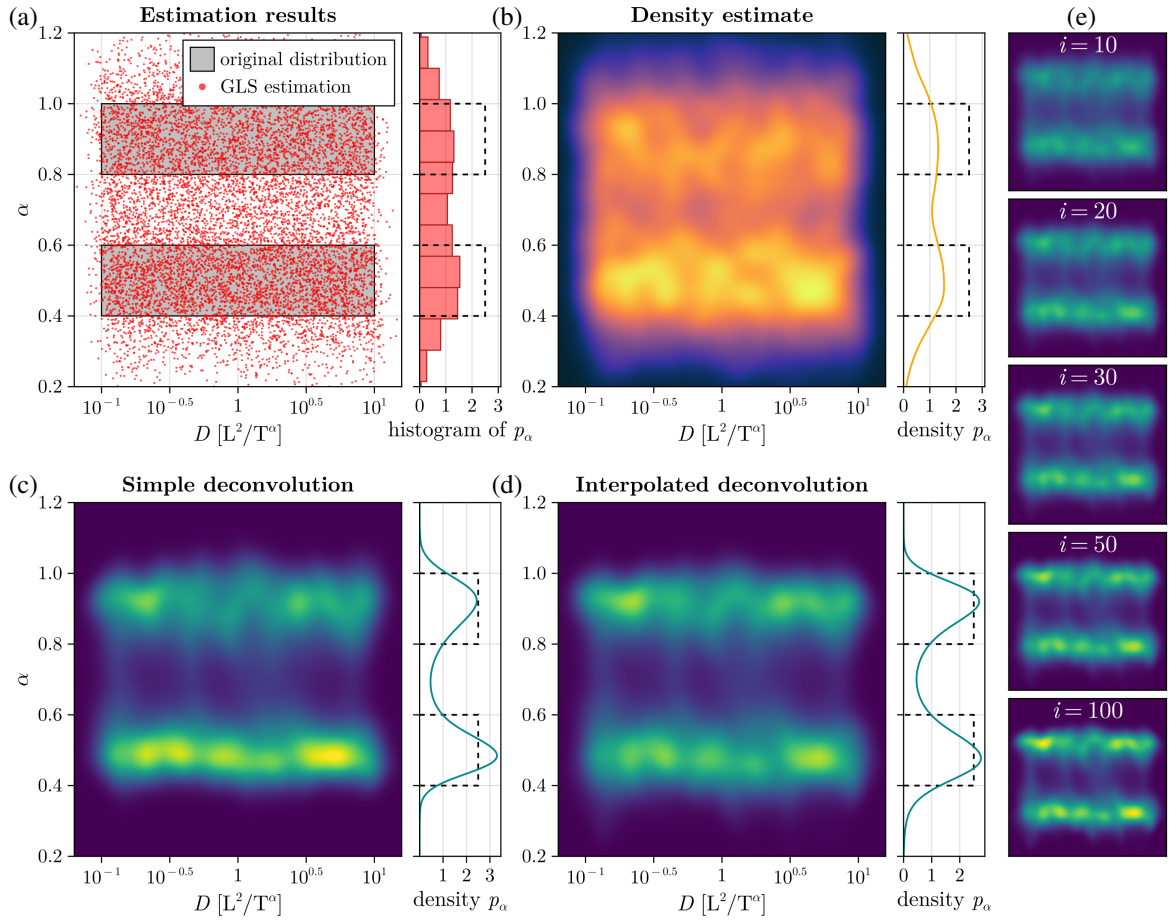

It is also worth checking what one can expect for a
more complicated original distribution. To this end, we draw parameters
of $10^4$ FBM trajectories from two rectangle distributions, as shown in
Fig.~\ref{fig:deconTest}. The analysis of the scatterplot of the TA-MSD fit, its
histogram or kernel density estimate, shows only a slight separation between
the two subpopulations. Using a deconvolution reveals much more of the true
structure of the original data. This is also proven by the $50\%$ decrease of
the $L^2$ distance between the original PDF and its estimate. The effects of
using a simple or interpolated deconvolution are mostly similar, but it can be
noticed that the simple deconvolution slightly overestimates small $\alpha$
values and underestimates larger ones, which is consistent with the error
covariance increasing with $\alpha$, see the bottom row of
Fig.~\ref{fig:deconTest}. In contrast, the interpolated deconvolution is
balanced for all $\alpha$.

\review{ In Fig.~\ref{fig:deconTest} the deconvolution revealed the main true features of the PDF, but also exhibits some smaller artefacts caused by the kernel density estimate, which are more likely to appear for more complicated original
distributions. These erroneous details become more sharp as the Richardson-Lucy iteration number increases, which is important to keep in mind when interpreting the result of deconvolution.}

\review{The quality of the final result strongly depends on quality of the original PDF estimate. This in turn depends on using a method appropriate for a particular dataset, which is beyond the scope of this work. However, if a kernel density estimate was used as the estimate, the deconvolution procedure can be further enhanced. The kernel density estimate works by blurring the empirical distribution of the sample, so one can deconvolve both the parameter estimation blur and  kernel estimate blur together by adding the corresponding covariance matrices and then using the corrected and wider $p^{\text{err+kde}}$. This is a procedure which was used in Fig.~10. }

\subsubsection{Parametric deconvolution}
\label{s:pDeconv}

For the parametric deconvolution we can use the fact that a convolution with
Gaussian noise is simple, for the source is a Gaussian itself. For a Gaussian
distribution with mean $\boldsymbol{\mu}$ and covariance $\Sigma^\mathrm{true}$,
we have $p^\mathrm{obs}=p^\mathrm{true}_{\mathcal{N}(\boldsymbol{\mu},\Sigma
	^\mathrm{true})}*p^\mathrm{err}_{\mathcal{N}(0,\Sigma^\mathrm{err})}=p_{\mathcal{
		N}(\boldsymbol{\mu},\Sigma^\mathrm{true}+\Sigma^\mathrm{err})}$. Thus, we can
reconstruct the original distribution as a normal distribution, $\mathcal{N}
(\boldsymbol{\mu},\Sigma^\mathrm{obs}-\Sigma^\mathrm{err})$. This is possible
only if $\Sigma^\mathrm{obs}-\Sigma^\mathrm{err}$ is a positive definite matrix,
which is a natural requirement stating that the blurred distribution has higher
variance along any axis than the errors. For two-dimensional matrices this
condition is easy to check.

A Gaussian distribution of $(\log D,\alpha)$ is a very limited case, but it can
be extended to any Gaussian mixture, that is, for any PDF of the form $p=\sum_k
w_kp_{\mathcal{N}(\boldsymbol{\mu}_k,\Sigma_k)}$ ($\sum_kw_k=1$) a deconvolution
would mean replacing each $\Sigma_k$ by $\Sigma_k-\Sigma^\mathrm{err}$.
Importantly, this procedure can be localized---for each mode we can use the
value $\Sigma^\mathrm{err}$ around $\boldsymbol{\mu}_k$.

\section{Summary of the statistical procedure}
\label{s:statSum}

For ease of use, we here show how a typical data analysis may be accomplished,
step by step.\\

\noindent
Input: A set of anomalous diffusion trajectories $X_k$, each can have its own
individual diffusivity $D$ and anomalous exponent $\alpha$.\\

\noindent Output: Estimated values of diffusivity $\widehat{D}$ and anomalous
diffusion exponent $\widehat{\alpha}$ and their errors. Optionally, also the
estimate of their joint distribution in which the influence of the estimation
errors are removed.\\

\noindent
Regression procedure:
\begin{enumerate}
	\item Calculate the TA-MSD as defined in Eq.~2.
	\item Calculate the log TA-MSD for a given logarithm base $b$.
	\item Obtain a preliminary estimate of $\alpha$ for each trajectory. You can
	use:
	\begin{itemize}
		\item[(i)] Some estimate known from previous studies. For example, if we know
		that we are dealing with a heterogeneous anomalous diffusion systems with mostly
		$0.6\le\alpha\le0.8$, then the global estimate $\alpha\approx0.7$ can be used.
		\item[(ii)] Trajectory-wise estimate from OLS, formula \ref{eq:OLS}. Note
		that only a few first values of TA-MSD should be used because, for larger
		windows, OLS quickly loses accuracy.
	\end{itemize}
	\item Calculate the log-error covariance matrix $\Sigma$ for each trajectory
	using Eq.~\ref{eq:sigma}. For this, you need to estimate the trajectory
	covariance, which depends on the model and the preliminary value of $\alpha$.
	If many trajectories need to be precessed, it might be good to use relation
	\ref{eq:reduced} to speed up computations.
	\item Remove the bias from log TA-MSD by adding $\ln b\Sigma_{k,k}/2$.
	\item Obtain the refined GLS estimate \ref{eq:gls}. In this step for the optimal result you can use
	all values of the TA-MSD.
	\item Estimate errors of $\log D$ and $\alpha$ from Eq.~\ref{eq:glscov} using
	the estimated pair $(\log \widehat{D},\widehat{\alpha})$ as input for the
	trajectory covariance matrix.
	\item If the distribution of $D$ instead of $\log D$ is needed, use the formulas
	from Sec.~\ref{s:move} to obtain an unbiased estimator of $D$ and estimate its
	error.
\end{enumerate}
In the presence of additive noise the procedure is slightly modified: in step 2
you first subtract twice the error variance, and in step 4 you use the modified
error covariance, which requires preliminary estimates of $\alpha$ and $D/
\sigma^2$, see Sec.~\ref{s:noise}.\\

\noindent
If one is interested in analyzing the distribution of $D$ and $\alpha$, in
addition:
\begin{enumerate}
	\setcounter{enumi}{8}
	\item Obtain estimates from the splitted MSD estimators \ref{eq:msdSplit} and
	check the correlation of $D$ and $\alpha$. It should have a lower artificially
	introduced correlation than the previous method at the cost of overall
	efficiency in estimating $D$ and $\alpha$.
	\item Estimate the joint PDF of $D$ and $\alpha$ using, e.g., histogram or
	kernel density estimator.
	\item Reconstruct the original PDF by removing the influence of statistical errors using the methods from Sec.~\ref{s:npDeconv} or
	\ref{s:pDeconv}. The choice depends on how much we know about the structure of
	this PDF. If we can divide it into some known, roughly Gaussian, subpopulations
	Sec.~\ref{s:pDeconv} should be better, otherwise use Sec.~\ref{s:npDeconv}.
\end{enumerate}

Implementations of this procedure in Julia and Python programming languages are available at \url{https://github.com/jaksle/anDiffReg}.


\end{document}